# Exploring ultra-slow-roll inflation in composite pseudo-Nambu–Goldstone boson models: implications for primordial black holes and gravitational waves

**Marco Merchand**[a]

Kobayashi-Maskawa Institute for the Origin of Particles and the Universe, Nagoya University, Tokai National Higher Education and Research System, Furo-cho, Chikusa-ku, Nagoya 464-8602, Japan

Received: 31 March 2026 / Accepted: 21 June 2026

**Abstract** We study inflation driven by a scalar potential arising from composite-sector dynamics, inspired by generalized composite Higgs models. The introduction of a non-minimal coupling, possessing the same functional form as the potential, induces a flattening at large field values that enables successful inflation. We analyze the conditions for ultra-slow-roll inflation, which leads to enhanced curvature perturbations, by combining analytical criteria near the inflection point with comprehensive numerical scans of the parameter space. The region consistent with cosmic microwave background constraints and yielding approximately $N_e \approx 55-60$ e-folds also predicts primordial black holes with masses in the range $10^3-10^5$ g. Although such ultra-light primordial black holes are typically expected to have evaporated, recent proposals invoking evaporation suppression via memory-burden effects could allow their survival as viable dark matter candidates. Under this assumption, the predicted gravitational wave signal lies in a frequency range currently inaccessible to any existing or proposed detectors. Although no experimental proposals presently reach this frequency band, our results provide strong motivation to push the frontiers of gravitational wave detection towards these unexplored high-frequency regimes.

## Contents

[a] e-mail: merchand.medina.marco.antonio.n1@f.mail.nagoya-u.ac.jp (corresponding author)

## 1 Introduction

Cosmic inflation, originally introduced to solve the horizon and flatness problems of the early universe [1], requires a significant amount of flatness of the inflaton potential to sustain a sufficient period of accelerated expansion. Realizing such a flat potential that is stable under quantum corrections is a central challenge in model-building. In the absence of an organizing symmetry, generic inflationary potentials are unprotected against radiative and gravitational corrections, raising concerns about their naturalness and predictivity.

A compelling resolution to this dilemma is provided by identifying the inflaton as a pseudo-Nambu–Goldstone boson (pNGB) [2–4]. In this scenario, the inflaton inherits an approximate continuous shift symmetry, $\phi \to \phi + C$, descending from the spontaneous breaking of a global symmetry $G \to H$. The leading-order potential for the Goldstone mode is strictly flat, and any explicit breaking (e.g., via gauge, Yukawa, or gravity-induced effects) only generates a potential that is radiatively stable, protected by the residual symmetry.

The archetypal model is “Natural Inflation”, where the inflaton is an axion and the potential is generated by instanton effects,

$$V(\phi) = \Lambda^4 \left[1 + \cos\left(\frac{\phi}{f}\right)\right], \tag{1.1}$$

with $f$ the decay constant. However, achieving consistency with current cosmological data typically requires a super-Planckian decay constant $f$, which poses challenges for the validity of the effective field theory, as Planck-suppressed higher-dimensional operators could destabilize the flatness of the potential. Additionally, the model suffers from large field excursions, which may further complicate its theoretical control. Despite these issues, the theory of natural inflation, along with various extensions and amendments, has been extensively studied in the literature [5–8] and has recently experienced renewed interest [9–11].

A particularly intriguing phenomenon in single-field inflationary models with suitably flat potentials is the emergence of transient ultra-slow-roll (USR) phases [12–15]. Near-inflection points in the inflaton potential can temporarily flatten the slope, dramatically decelerating the inflaton and amplifying the primordial curvature perturbations by several orders of magnitude over small scales. Such amplified perturbations can gravitationally collapse upon horizon reentry to produce primordial black holes (PBHs) [16], which are prime candidates for dark matter [17].

In their seminal works, Ballesteros et al. [14,18] have emphasized the importance of performing an exact computation of the primordial curvature perturbation to accurately predict PBH abundances. Using the Mukhanov–Sasaki formalism, they provide the most diligent and precise calculation of the power spectrum, particularly near the formation of the peak, where the commonly used slow-roll approximation fails severely. The slow-roll methods can underestimate the peak amplitude by several orders of magnitude, leading to significant errors in estimating PBH production. Furthermore, these models exhibit a well-known strong fine-tuning of the inflationary potential parameters, which is necessary to achieve the required enhancement in the power spectrum for viable PBH dark matter production [19].

While USR single-field inflationary models can produce the enhanced power spectrum required for PBH dark matter in the asteroid mass range, they face a notorious difficulty in fitting the observed spectral index $n_s$ of primordial fluctuations from cosmic microwave background (CMB) data, yielding values that are somewhat lower than the measurements, typically around $n_s \lesssim 0.95$, with the new combined Atacama Cosmology Telescope (ACT) results [20,21], exacerbating the tension, making it an open problem to reconcile PBH production with precision cosmology constraints effectively. This challenge motivates considering slight potential modifications or alternative cosmologies to accommodate both sets of observations consistently.

The expected mass range for PBH dark matter is tightly constrained: microlensing and dynamical constraints set an upper bound, while the lower bound arises from the standard Hawking evaporation process [22,23], which predicts that PBHs lighter than approximately $10^{14}$ g have evaporated by the current epoch (see Refs. [16,17,24–27] for comprehensive reviews). Importantly, this semiclassical evaporation assumes a self-similarity in the black hole evolution, whereby an old PBH is indistinguishable from a newly formed one of the same mass.

Recent theoretical advances challenge the standard assumption of classical Hawking evaporation for PBHs by incorporating quantum information effects that induce the so-called memory-burden phenomenon [28–30]. This effect significantly slows the evaporation rate once a PBH has lost roughly half its initial mass, effectively stabilizing ultra-light PBHs and enabling their possible survival until today. If the memory-burden sets in sufficiently early, possibly soon after formation [31], it opens a previously unexplored window for ultralight PBHs to constitute a substantial fraction of dark matter.

The evaporation rate incorporating the memory-burden effect is modeled by

$$\frac{dM}{dt} = -\frac{\gamma}{M^2}\left[\Theta M - qM_{\rm PBH,i} + \frac{1}{S^k}\,\Theta(qM_{\rm PBH,i} - M)\right], \tag{1.2}$$

where $\gamma$ encompasses particle physics parameters, $k$ quantifies the entropy suppression exponent, $S = 4\pi G M^2$ is the black hole entropy, $M_{\rm PBH,i}$ is the initial PBH mass and $\Theta$ is the Heaviside function selecting the evaporation regime. The parameter $q$ denotes the fractional mass threshold separating the two evaporation regimes. This formulation describes a sharp or gradual transition that significantly prolongs PBH lifetimes. The extent of this lifetime extension depends sensitively on $q$ and $k$, both of which remain to be precisely constrained by theory or observation. This is equivalent to the piecewise form adopted in e.g. [31,32], with the Hawking (memory-burden) phase active for $M > qM_{\rm PBH,i}$ ($M \leq qM_{\rm PBH,i}$).

Reference [31] examines the transition from standard semi-classical black hole evaporation to the memory-burdened phase using a smooth ansatz based on a hyperbolic tangent function. They find that if this transition is gradual, PBHs lighter than about $4 \times 10^{16}$ g cannot make up a significant fraction of dark matter for a wide range of fractional mass parameters $q = 0.2 - 0.9$. However, an almost instantaneous onset of memory-burden ($1 - q \lesssim 10^{-10}$) with a suppression exponent $k = 2$ opens a previously forbidden window, allowing PBHs with masses as low as $10^4$ g to account for all dark matter, supporting the viability of *ultra-light* PBHs.

A complementary study [33] employs a prototype master mode Hamiltonian focusing on the transition width and shows that the gradual nature of the transition significantly impacts observational bounds on light PBHs – especially through neutrino fluxes – independently of the final suppression exponent. While broadly consistent with [31], their different parametrizations highlight complementary phenomenological aspects.

Other works studying constraints on the memory-burdened PBHs from Big Bang Nucleosynthesis, CMB [34–36], and high-energy neutrinos [37,38] typically assume $q = 0.5$. These works analyze the two-dimensional parameter space of $M_{\rm PBH}$ and entropy suppression exponent $k$. Notably, Ref. [39] finds that for $k = 2$ and $q = 0.5$, ultralight PBHs can comprise all of dark matter only within a narrow mass window $10^9 \lesssim M_{\rm PBH}/{\rm g} \lesssim 10^{10}$, indicating the necessity of an early memory-burden phase onset. Extending these analyses to larger $k$ values remains an open question.

In summary, ultra-light PBHs can serve as viable dark matter candidates if the memory-burden effect activates early during their evaporation or if the entropy suppression exponent $k$ attains values larger than approximately $k \gtrsim 2.5$. Furthermore, even disregarding the existence of a memory-burden phase and the role of these PBHs as dark matter, the identified ultra-light PBH mass ranges remain of significant cosmological interest. These mass ranges can substantially contribute to the generation of the baryon asymmetry and dark matter of the Universe through mechanisms involving Hawking radiation and entropy injection. For an incomplete list of relevant references, see [40–44]. This connection motivates further investigation of these ultra-light PBHs beyond the conventional dark matter paradigm.

The amplification of primordial perturbations during USR phases also induces a stochastic background of gravitational waves (GW) through second-order effects [45,46]. The peak frequency of these induced GWs is directly correlated with the PBH mass, establishing a one-to-one correspondence between the two. For PBHs within the preferred asteroid mass range ($10^{17}{\rm g} \lesssim M_{\rm PBH} \lesssim 10^{21}{\rm g}$), the GW signals peak at frequencies accessible to the Laser Interferometer Space Antenna (LISA) [47–49]. Conversely, PBHs formed at smaller scales produce lighter masses and correspondingly higher-frequency GWs – on the order of MHz-GHz – which may be detectable by next-generation advanced interferometers and resonant cavity experiments [50–53].

Trigonometric potentials, such as those in natural inflation and its variants, provide a natural framework to realize these inflationary features. Such potentials have been extensively utilized to model large primordial curvature perturbations and USR dynamics in various contexts, including hybrid inflation [54], composite hybrid inflation [55,56], string-inspired axionic potentials with localized bumps [57–59], natural inflation with non-trivial kinetic terms [60,61], and $\alpha$-attractor models [62]. While some rely on modifications or additions to purely trigonometric functions, such as logarithmic or hyperbolic terms, here we focus on a purely trigonometric potential form inspired by composite Higgs models.[1]

Motivated by recent developments in composite inflaton scenarios [55], we propose a novel single-field inflationary model where the inflaton is a pNGB arising from a composite sector (as in the minimal composite Higgs scenario of [66]). The inflaton potential features an extended trigonometric structure, and the theory incorporates a periodic non-minimal coupling to gravity as introduced in [67]. Unlike prior models which often suffer from trans-Planckian decay constants or rely on non-canonical kinetic terms, our construction leverages the composite origin to generate natural steepening and flattening regions in the potential. This structure facilitates an ultra-slow roll phase conducive to producing significant primordial curvature perturbations, PBHs, and their associated stochastic GW signatures.

Unlike much of the existing literature, where authors typically fix a target PBH mass range and then tune or construct an inflationary potential to realize this mass, our approach is fundamentally more exploratory. We do not presuppose the final PBH mass outcome; instead, we systematically investigate the full parameter space of the model, guided primarily by the necessary conditions for ultra-slow roll inflation. This allows us to uncover the range of naturally emerging inflationary scenarios and their associated phenomenology without bias towards any specific PBH mass scale. To the best of our knowledge, this novel inflationary model inspired by composite-sector dynamics has not previously been examined in the context of PBH formation and GW signatures.[2]

By performing a comprehensive and generic scan over the parameter space of our model, rather than fixing a PBH mass scale *a priori*, we find that the natural outcome includes the production of ultra-light PBHs. Remarkably, these PBHs fall within a mass regime where, in order to account for their viability as dark matter candidates, one must invoke or assume that they experienced a memory-burdened phase affecting their evaporation and stability. This is not to say that the obtained PBH mass inherently guarantees the memory-burden effect, but rather that such ultra-light masses necessitate this assumption to remain viable dark matter candidates. This contrast with the common paradigm underscores the predictive power and distinct phenomenology of our model, opening new avenues for observational tests and theoretical developments.

[1] Model cases which do not employ trigonometric functions are more abundant, some examples include Higgs inflation [63], multi-field inflation models [64], Einstein–Gauss Bonnet gravity [65].

[2] Papers that study ultra-light PBH masses without assuming a particular inflationary model include [32,68,69]. Other works that investigate PBHs from curvature perturbations in a model agnostic way include [70].

The rest of the paper is organized as follows. In Sect. 2, we introduce the composite inflaton potential model, detailing its main properties and origins, along with the full Jordan frame action incorporating non-minimal coupling to gravity. Section 3 is devoted to the analysis of the USR phase, where we describe the conditions for USR inflation and present our benchmarking approach to extract the relevant parameter space of the model. Readers interested primarily in the phenomenological results may skip this section and proceed directly to Sect. 4, which contains our main findings, including predictions for primordial black hole production and gravitational wave signatures. Finally, in Sect. 5, we provide concluding remarks and outline possible directions for future research.

The calculation of the power spectrum and PBH abundance follows the well-established procedures outlined in [14], which we first reproduced as a cross-validation of our numerical implementation. The GW spectrum calculation closely follows the methodology detailed in [45]. For completeness, we present the most relevant formulas in Appendix, while we encourage the reader to consult these references for a more comprehensive explanation of the underlying calculations.

## 2 Composite Higgs mechanism: a robust origin for pNGB potentials

In the framework of the Callan–Coleman–Wess–Zumino formalism, the Goldstone bosons associated with the spontaneous symmetry breaking of a global symmetry group $G$ down to a subgroup $H$ are parameterized by the coset space $G/H$. The Goldstone fields $\pi^{\hat{a}}(x)$, corresponding to the broken generators $T^{\hat{a}}$ of $G/H$, are assembled into a non-linear sigma model matrix via the exponential map [71,72]

$$\Sigma(x) = \Sigma_0 \exp\left(i\frac{\pi^{\hat{a}}(x)T^{\hat{a}}}{f}\right) = \frac{1}{\phi}\sin\left(\frac{\phi}{f}\right)\begin{pmatrix}\pi_1\\ \vdots \\ \pi_N \\ \phi\cot(\phi/f)\end{pmatrix}, \quad \phi \equiv \sqrt{\pi^a\pi^a}, \tag{2.1}$$

where $\Sigma_0 = (0, ..., 1)$ encodes the spontaneous symmetry breaking direction and $f$ is the symmetry breaking scale. This construction ensures that the theory contains only derivatively coupled Goldstone bosons, reflecting the shift symmetry and guaranteeing the absence of a potential at the classical level. The matrix $\Sigma(x)$ transforms non-linearly under $G$, implementing the Goldstone nature of the fields and setting the stage for constructing invariant effective Lagrangians.

A well-motivated scenario identifies the inflaton as a pNGB arising from the confinement dynamics of a strongly interacting gauge theory. Analogous to composite Higgs models, the scalar potential for this field is absent at tree level and generated radiatively from loop corrections via the Coleman-Weinberg mechanism and augmented by explicit symmetry-breaking mass terms. Thus, the inflaton emerges as a pNGB from the spontaneous breaking of a global symmetry $G \to H$ [66]. The pNGB's effective potential is thus induced by gauge and fermion loops, typically taking the form [66]:

$$V_{\rm inf}(\phi) = \Lambda^4 P(\phi), \tag{2.2}$$

with[3]

$$P(\phi) = a_1\left[1 - \cos\left(\frac{\phi}{f}\right)\right] - a_2\sin^2\left(\frac{\phi}{f}\right) + a_3\sin^4\left(\frac{\phi}{f}\right), \tag{2.3}$$

where the $a_i$ are calculable coefficients determined by the underlying strong dynamics, gauge couplings, or couplings to explicit symmetry-breaking operators [5,6,66]. In particular, the first term encapsulates explicit symmetry breaking from a mass term while the last two terms appear from radiative loop effects. In the above expression, the very first term which is field independent is a cosmological constant introduced by hand such that the potential vanishes at the origin of field space, i.e., $V_{\rm inf}(0) = 0$.

We emphasize that such a structure is not postulated ad-hoc: it is a generic prediction of composite pNGB models, including the Minimal Composite Higgs Model [66], and its inflationary counterparts as studied in Refs. [5,6]. Importantly, for suitable choices of $a_i$ and $f$, this potential naturally exhibits regions of steepening and flattening – critical for realizing both slow-roll and ultra-slow-roll phases as we will cover in the following subsections.

### 2.1 Non-minimal coupling and canonical field space

In realistic cosmological settings, it is natural to generalize the minimal coupling of the pNGB inflaton to include possible non-minimal couplings to gravity. Thus, the generic Jordan-frame action reads

[3] Similar potentials with multiple sinusoidal terms have been studied previously, notably in the multi-natural inflation scenario [73], which explores implications for spectral predictions and field ranges. While our model differs in specifics, it shares conceptual similarities that merit further investigation.

$$S_{\text{Jordan}} = \int d^4x\sqrt{-g}\left[\frac{M_{\text{Pl}}^2}{2}F(\phi)R + \frac{1}{2}g_{ab}(\pi)\partial_\mu\pi^a\partial^\mu\pi^b - V_{\text{inf}}(\phi)\right], \tag{2.4}$$

where $M_{\text{Pl}} = 2.8435 \times 10^{18}$ GeV is the reduced Planck mass.

We introduce a non-minimal coupling to gravity as follows

$$F(\phi) \equiv 1 + \alpha P(\phi), \tag{2.5}$$

where the dimensionless parameter $\alpha$ controls the strength of the non-minimal interaction, and $P(\phi)$ is the trigonometric potential function characterizing the composite-sector dynamics, see Eq. (2.3). This represents the simplest non-minimal coupling consistent with the discrete shift symmetry of the inflaton's radial mode. Moreover, it is proportional to the potential itself and ensures that standard Einstein gravity is recovered at the global minimum of the potential.

The UV origin of such a periodic non-minimal coupling may be traced back to operators in the underlying composite theory, for instance, terms coupling the fermionic constituents to curvature, which upon condensation induce trigonometric structures similar to those appearing in the potential. A particularly compelling illustration of this mechanism is provided in recent work by Salvio [9], where the authors investigate a multifield inflationary scenario involving a pNGB and a scalaron arising from a $R^2$ term in the gravitational action. In their framework, a non-minimal coupling between the pNGB and gravity naturally emerges from quantum gravity effects and the underlying strong dynamics that generate the composite sector. Specifically, non-renormalizable interactions between the composite sector fermions and curvature can, after spontaneous chiral symmetry breaking and condensation, induce effective operators involving the Ricci scalar multiplied by trigonometric functions of the inflaton field.

Earlier work by Ferreira et al. [67]. also considered a periodic non-minimal coupling proportional to the inflaton potential, respecting the discrete shift symmetry, and showed how such a coupling improves agreement with cosmological data. To the best of our knowledge, their study was the first to introduce this specific form of non-minimal coupling in the context of natural inflation, although it did not provide a microscopic derivation or the physical origin of such an interaction.

Despite these advances, a rigorous microscopic derivation of periodic non-minimal coupling from a fundamental UV-complete theory remains an open problem. The existence of such effective operators is, however, well motivated within the context of composite Higgs or axion-like models coupled to gravity. Further investigation into the detailed UV-completion and the exact form of these induced operators is warranted and constitutes an important avenue for future research.

For the purposes of this work, we focus exclusively on the radial (or modulus) direction of the scalar manifold, which corresponds to the inflaton field driving the background cosmological dynamics. The full spontaneous symmetry breaking $G \to H$ generates multiple true Goldstone bosons associated with the broken generators, populating a multi-dimensional field space. However, these Goldstone modes are derivatively coupled and essentially massless at the classical level. In fact, the equations of motion simplify to the canonical form only for the radial direction, effectively decoupling the Goldstone directions from the dynamics. This allows us to consistently truncate the full multi-field system to a single-field effective description focused on the radial mode. Correspondingly, the field space metric simplifies significantly, eliminating kinetic mixing terms that involve the Goldstone fields in our analysis.

To analyze the inflationary dynamics, we perform a Weyl transformation via $g_{\mu\nu} \to F(\phi)^{-1}g_{\mu\nu}$ such that the action for the canonically normalized inflaton field in the Einstein frame reads:

$$S_{\text{Einstein}} = \int d^4x\sqrt{-g}\left[\frac{M_{\text{Pl}}^2}{2}R + \frac{1}{2}\partial_\mu\chi\partial^\mu\chi - U(\chi)\right], \tag{2.6}$$

with $U(\chi) = U(\phi(\chi))$ the effective inflaton potential, see Eq. (A.3).

For the predictions presented in this work, we exclusively use the canonically normalized inflaton field, $\chi$. However, in our benchmarking procedure, we adopt a hybrid approach wherein analytical expressions are more conveniently handled using the potential expressed as a function of the original field, $U(\phi)$, rather than the canonically normalized potential $U(\chi)$. For completeness and reader convenience, we provide a detailed derivation of the relevant formulas – including the Weyl transformation, kinetic function, and the canonical field redefinition – in Appendix A.

## 3 Ultra-slow-roll phase

USR inflation arises near an inflection point in the potential, where the slope becomes extremely flat and the inflaton acceleration increases substantially, violating the second slow-roll condition, i.e. $\eta_H \geq 3$ see Eq. (B.1), thus the reason for the name USR phase, see Refs. [74,75] for comprehensive reviews. This transient regime can amplify primordial curvature perturbations by several orders of magnitude over a limited range of scales, a phenomenon crucial for efficient PBH production. Key requirements for USR include the existence of an approximate inflection point and sufficiently prolonged flatness in the potential. Notably, the seminal works by Ivanov et al. [12], Germani and Prokopec [13], Bellido and Morales [15], and more recent precise analyses by Ballesteros et al. [14,18] have rigorously elucidated the dynamics of ultra-slow-roll inflation and its impact on PBH formation. These studies emphasize the critical necessity of employing exact numerical computations of curvature perturbations, as the usual slow-roll approximations fail to capture the enhanced power spectrum accurately during the USR phase. In the following section, we detail our benchmarking and parameter extraction algorithm, ensuring that all conditions for USR are satisfied within our composite pNGB framework.

### 3.1 Parameter extraction: benchmarking algorithm

Throughout this work, we focus on inflaton potentials that roll towards the origin in field space, following the principle that flattening occurs at large field values due to the non-minimal coupling.[4] Given that the potential is even and exhibits a periodicity of $2\pi$, we restrict our analysis to the field space region $\phi \in [0, \pi f]$, ensuring that the entire inflationary dynamics – from the beginning to the end of inflation – are encompassed within this interval.

Given these assumptions, the inflaton must begin its evolution near the field value $\phi = \pi f$, slowly rolling down the potential towards smaller values. Upon approaching a near-inflection point, the inflaton enters an USR phase, during which it spends several e-folds lingering in the nearly flat region. Eventually, the field exits this zone and resumes slow-roll descent, concluding inflation when the slow-roll parameter $\epsilon_H$ reaches unity, as defined in Eq. (C.4). A natural requirement for this dynamics is that the potential remains monotonically increasing for $\phi < \pi f$, except near the inflection point where flattening occurs.

Following this desired behavior, we summarize the essential constraints to be imposed on the scalar potential:

1. $E_1 \equiv U(0) = 0,\quad$ (Global minimum),

2. $E_2 \equiv U(\pi f) > 0,\quad$ (Potential positive at boundary), (3.1)

**At the inflection point $\phi_0$ :**

3. $E_3 \equiv \left.\dfrac{dU}{d\phi}\right|_{\phi_0} \approx 0,\quad$ (Zero slope), (3.2)

4. $E_4 \equiv \left.\dfrac{d^2U}{d\phi^2}\right|_{\phi_0} \approx 0,\quad$ (Zero curvature), (3.3)

**At the boundary $\phi = \pi f$ :**

5. $E_5 \equiv \left.\dfrac{dU}{d\phi}\right|_{\pi f} \approx 0,$ (Flatness at boundary),

6. $E_6 \equiv \left.\dfrac{d^2U}{d\phi^2}\right|_{\pi f} \approx 0,$ (Curvature vanishes at boundary), (3.4)

7. CMB constraints as per Eq. (B.4).

Notice that we introduced the inflection point $\phi_0$ as a free parameter. Conditions 1–6 are evaluated using the original Jordan frame field $\phi$, which results in simpler expressions and allows for faster numerical implementation compared to the canonically normalized field. However, condition 7 is evaluated using the canonically normalized inflaton field, $\chi_*$, which denotes the field value at the time when CMB scales, corresponding to the pivot scale $k = 0.05\,\mathrm{Mpc}^{-1}$, exited the horizon, based on the most recent measurements. Practically, $\chi_*$ is obtained by solving for a desired value of the spectral index $n_s$ within the experimentally allowed $3\sigma$ range, and subsequently verifying that the other observables, such as the tensor-to-scalar ratio $r(\chi_*)$ and running $\alpha_s(\chi_*)$, satisfy the required constraints.

Notice that conditions 1 and 5 are trivially satisfied for our choice of potential parametrization. The other constraints can be written analytically, for example, to guarantee sufficient flatness near the potential maximum, condition 6, yields

$$(a_1 + 2\,a_2)(-1 + 2a_1\alpha) = 0, \tag{3.5}$$

furthermore, condition 2 yields

$$2a_1 > 0, \tag{3.6}$$

these conditions are supplemented with conditons 3 and 4 which depend on the inflection point location and can become cumbersome in some cases.

The potential is characterized by six free parameters: two dimensionful scales, $\Lambda$ and $f$, and four dimensionless parameters, $a_1$, $a_2$, $a_3$, and $\alpha$. Once a suitable potential shape is obtained, the overall scale $\Lambda$ is fixed by matching the observed amplitude of the power spectrum at the pivot scale, while the decay constant $f$ is chosen to be close to unity (in Planck units) to maintain naturalness without requiring large trans-Planckian values. This leaves four dimensionless parameters controlling the detailed shape. There are four non-trivial constraints on the potential; however, one introduces the inflection point $\phi_0$ as an additional free parameter. By expressing $\phi_0$ in terms of the remaining parameters, this reduces to three constraints for four effective dimensionless parameters. Consequently, one parameter remains free, which must be fine-tuned to generate large peaks in the curvature power spectrum necessary for primordial black hole formation. Remarkably, the minimal composite-Higgs mechanism naturally provides precisely the right number of parameters to construct a viable inflationary model featuring an ultra-slow-roll phase. Moreover, the requirement of three linearly independent functions to produce the requisite potential shape arises naturally in the composite Higgs framework, underscoring its suitability for this inflationary scenario.

[4] Interestingly, we have explicitly verified that inflaton rolling from left to right also yields viable solutions within our model, as demonstrated by a benchmark point obtained through dedicated analysis (though no extensive parameter scan was performed in this regime). This near-symmetric behavior arises from the intrinsic parity symmetry of the potential, which is symmetric between $\phi = 0$ and $\phi = \pi f$. Notably, to facilitate inflaton rolling from left to right consistently, one must invert the sign of the cosine term in the potential to ensure the potential vanishes at $\phi = \pi f$. Unlike conventional setups where the non-minimal coupling primarily flattens the potential at large field values guiding the inflaton's roll downhill (right to left), our model's unique multi-harmonic structure and composite origin naturally accommodate flattening on both "ends", enabling a reversed rolling direction with similar inflationary predictions.

In Fig. 1, we display our algorithm for selecting viable models consistent with an USR phase. We first choose (upper left corner in the figure) a random value for the inflection point $\phi_0 \in (0, \pi f)$, then solve conditions 2, 3, 4 and 6 (Eqs. (3.1)–(3.4)) in favor of $\alpha$, $a_1$, $a_2$, and $a_3$. The system usually has many solutions for these parameters, and we select only those cases in which the potential has the desired shape, namely, is flat close to the boundary of field space and has an inflection point at $\phi_0$, see Fig. 2. After checking if the CMB conditions, Eq. (B.4), are satisfied, we proceed to solve the equation of motion, Eq. (B.5), checking (1) if the number of e-folds is in a reasonable range, (2) that the inflaton does not get stuck in inflection point or (3) that inflation ends prematurely ($\epsilon_H > 1$) even before reaching the inflection point.[5] Verifying that all requirements are satisfied, we then evaluate the peak of the primordial curvature perturbation using the slow-roll approximation, Eq. (B.6). If the peak of the power spectrum raises above its value at the CMB scales, to about $10^{-8}$, we select the model to perform further numerical analysis so we can further increase the amplitude of the power spectrum. In order to achieve this, we chose to manually tune $a_3$ as shown differentiated with a red dashed line in Fig. 1. Once a benchmark it's selected we carry out the full numerical solution of the Mukhanov–Sasaki equation, see Appendix C.

We wish to clarify that, throughout our benchmarking procedure, the inflection point conditions $E_3$–$E_6$ are presented with expressions that are approximately zero; however, their actual numerical values are non-zero and are randomly initialized at the beginning of the algorithm. Setting these parameters exactly to zero would generically lead to the absence of viable solutions for the inflection point constraints, rendering the potential unsuitable for our purposes. Moreover, it is now well established in the literature – see, e.g., Refs. [14,59] – that the USR phase works optimally for a shallow local minimum or a near-inflection point, rather than for a mathematically exact inflection point. Here, the term "shallow" refers to a region of the potential where the curvature is small but positive, so the inflaton slows down substantially but is not trapped. This practical feature is crucial: while an exact inflection point corresponds to a vanishing slope and curvature, in realistic models and parameter searches, a shallow minimum provides sufficiently flat terrain for USR dynamics and amplification of primordial fluctuations, without inducing mathematical singularities or trapping the field.

In Fig. 2 we show the scalar potential for Benchmark 1 (see Table 1) in the Jordan and Einstein frames, respectively, highlighting that although the potential has already the right shape in the Jordan frame, the steepness around the increasing region between the near-inflection point and the top of the hill, is too large, causing inflation to stop before even reaching the inflection point. On the other hand, the non-minimal coupling induces in the Einstein frame the correct behavior and is thus critical to generate viable models. The green dots on top of the curve represent the inflaton trajectory at each time step of the numerical integration, with denser number of points representing slower inflaton velocities. We also show the inflaton velocity as it traverses down its scalar potential. First starting off with a small speed – dictated by slow-roll initial conditions, building up speed until it encounters the inflection point which brings it to a sharp slow-down and it continues rolling down once it overcomes the flat region until the slow-roll condition is finally violated and reheating can start.

To further illustrate the necessity[6] of the non-minimal coupling, Fig. 3 shows the predictions for the spectral index $n_s$ using the parameter values of Benchmark 1 in Table 1, compared with the case of vanishing non-minimal coupling. As evident, the model without the non-minimal coupling fails to achieve sufficiently large values of $n_s$ consistent with observations.

[5] We note that a transient violation of $\epsilon_U > 1$ before the USR phase need not be physically problematic. Although the potential becomes locally steep in this region, the inflaton's kinetic energy has not built up sufficiently to violate the accelerated-expansion condition, so the inflationary background with $\epsilon_H < 1$ is maintained throughout. After this transient, the inflaton rapidly enters the USR attractor, as expected. In practice, all benchmark points quoted in the results exhibit such a temporary region with $\epsilon_U > 1$, yet all maintain $\epsilon_H < 1$ until the onset of the USR phase.

[6] An intriguing possibility is that warm inflation [76], which includes dissipative effects sustaining a thermal bath during inflation, could allow for slow-roll dynamics on steeper Jordan-frame potentials. This might alleviate the need for a non-minimal coupling by naturally satisfying slow-roll conditions without extreme potential flattening. This remains speculative within our framework and warrants further exploration given the rich phenomenology of warm inflation models [77–79].

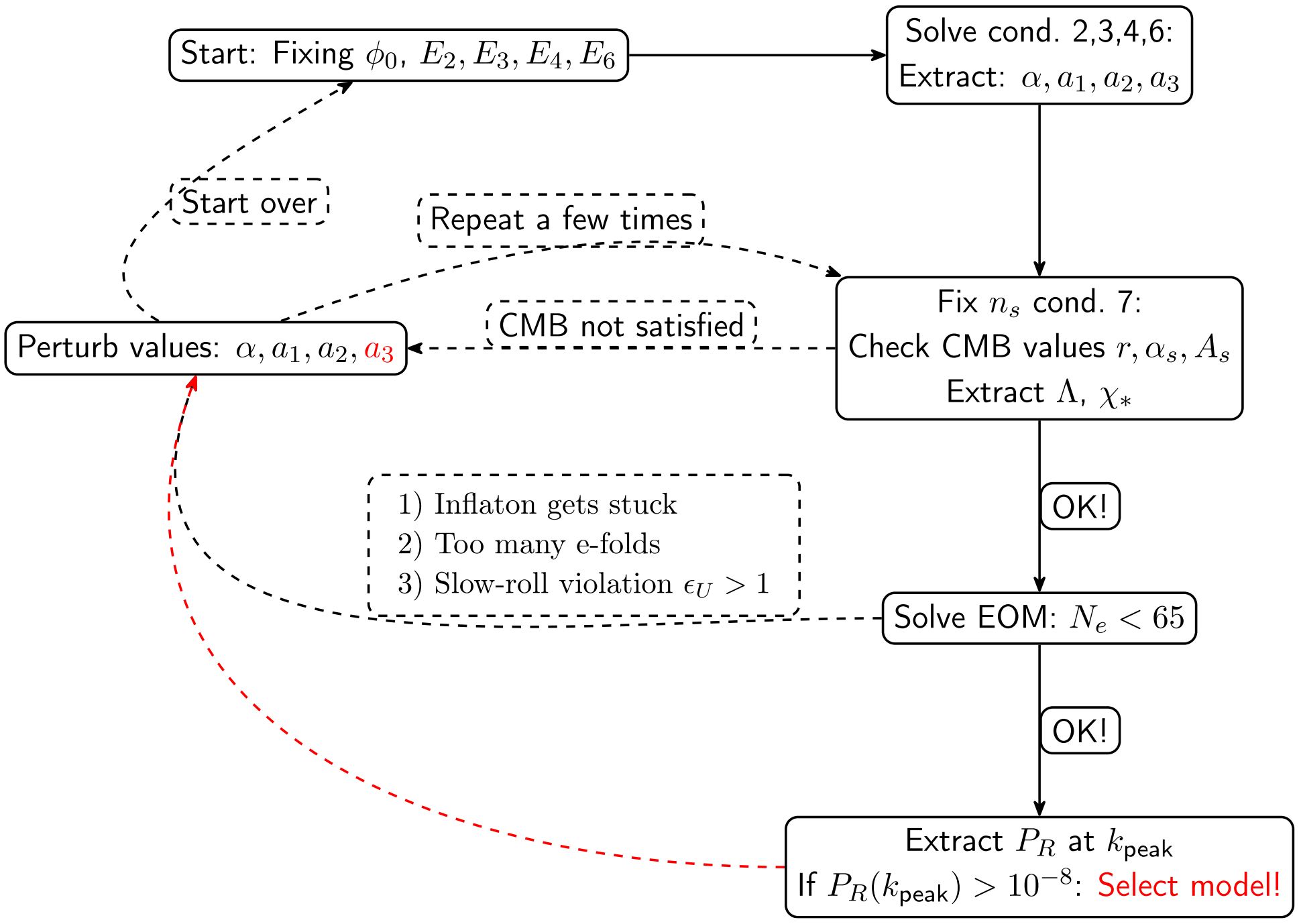

**Fig. 1** Schematic representation of the numerical algorithm used for benchmark extraction consistent with USR inflationary phase

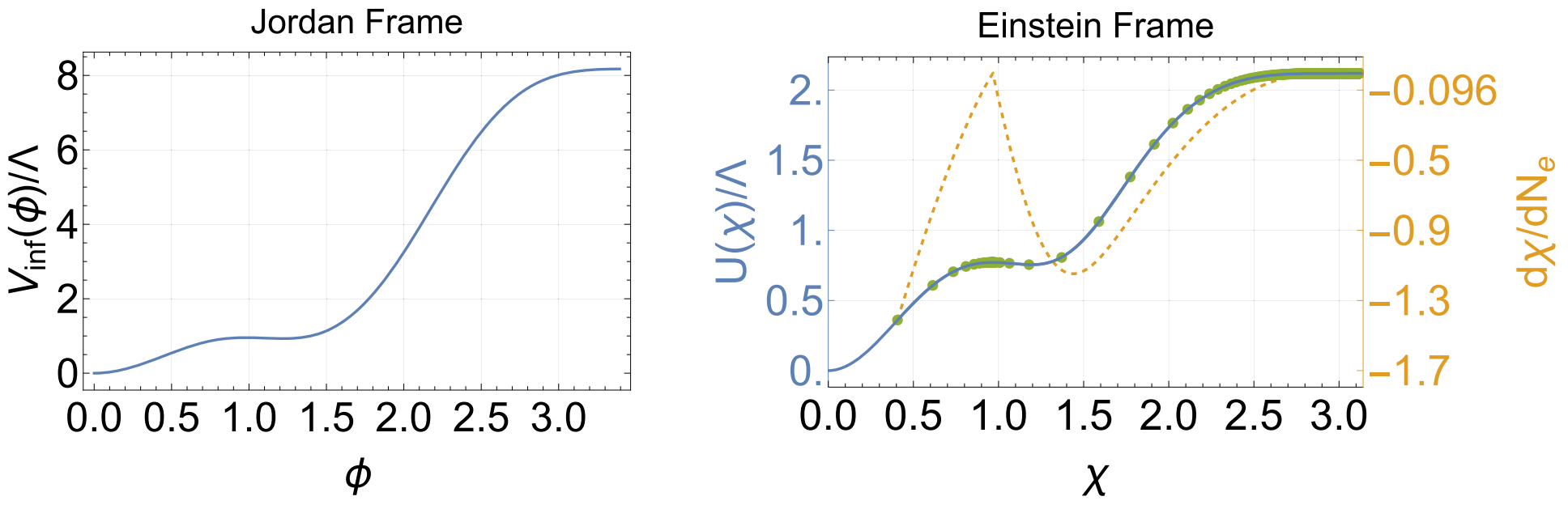

**Fig. 2** Illustrations corresponding to Benchmark 1 of Table 1. Left: Potential in the Jordan frame. Right: Potential in the Einstein frame. The green dots on the curve indicate the inflaton USR trajectory and the dashed-orange line the infaton velocity

**Table 1** Table of the most promising benchmark values. Dimensionful quantities are normalized to $M_{\text{Pl}}$

| Benchmark | $f$ | $\alpha$ | $a_1$ | $a_2$ | $a_3$ | $\frac{\phi_0}{f}/(\pi/3)$ | $\Lambda^4$ | $\chi_*$ | $N_{\text{e-folds}}$ |
|---|---|---|---|---|---|---|---|---|---|
| Benchmark 1 | 1.080 | 0.118 | 4.086 | −1.330 | −3.737 | 1.024 | $5.39 \times 10^{-13}$ | 2.949 | 59.561 |
| Benchmark 2 | 1.050 | 0.119 | 4.084 | −1.292 | −3.733 | 1.023 | $4.54 \times 10^{-13}$ | 2.872 | 58.623 |
| Benchmark 3 | 1.000 | 0.228 | 2.100 | −0.789 | −2.017 | 1.032 | $7.61 \times 10^{-13}$ | 2.768 | 59.987 |
| Benchmark 4 | 0.965 | 0.107 | 4.596 | −1.301 | −4.193 | 1.020 | $3.06 \times 10^{-13}$ | 2.696 | 57.030 |
| Benchmark 5 | 0.964 | 0.111 | 4.374 | −1.221 | −3.983 | 1.019 | $3.59 \times 10^{-13}$ | 2.714 | 56.815 |
| Benchmark 6 | 1.200 | 0.163 | 2.994 | −0.171 | −2.180 | 0.976 | $2.71 \times 10^{-13}$ | 3.363 | 91.705 |

**Table 2** Predicted cosmological parameters and PBH masses

| Benchmark | $n_s(\chi_*)$ | $r(\chi_*)$ | $\alpha_s(\chi_*)$ | $k_{\text{peak}}$ (Mpc$^{-1}$) | $M_{\text{PBH}}(g)$ |
|---|---|---|---|---|---|
| Benchmark 1 | 0.952 | $3.51 \times 10^{-5}$ | $-8.58 \times 10^{-4}$ | $5.09 \times 10^{20}$ | 18905.7 |
| Benchmark 2 | 0.952 | $2.92 \times 10^{-5}$ | $-1.21 \times 10^{-3}$ | $4.67 \times 10^{20}$ | 22422.7 |
| Benchmark 3 | 0.952 | $2.56 \times 10^{-5}$ | $-9.24 \times 10^{-4}$ | $3.47 \times 10^{21}$ | 406.204 |
| Benchmark 4 | 0.952 | $2.20 \times 10^{-5}$ | $-9.25 \times 10^{-4}$ | $8.58 \times 10^{20}$ | 6652.29 |
| Benchmark 5 | 0.952 | $2.48 \times 10^{-5}$ | $-1.10 \times 10^{-3}$ | $7.71 \times 10^{20}$ | 8243.49 |
| Benchmark 6 | 0.970 | $1.28 \times 10^{-5}$ | $-3.99 \times 10^{-4}$ | $7.27 \times 10^{34}$ | $9.28113 \times 10^{-25}$ |

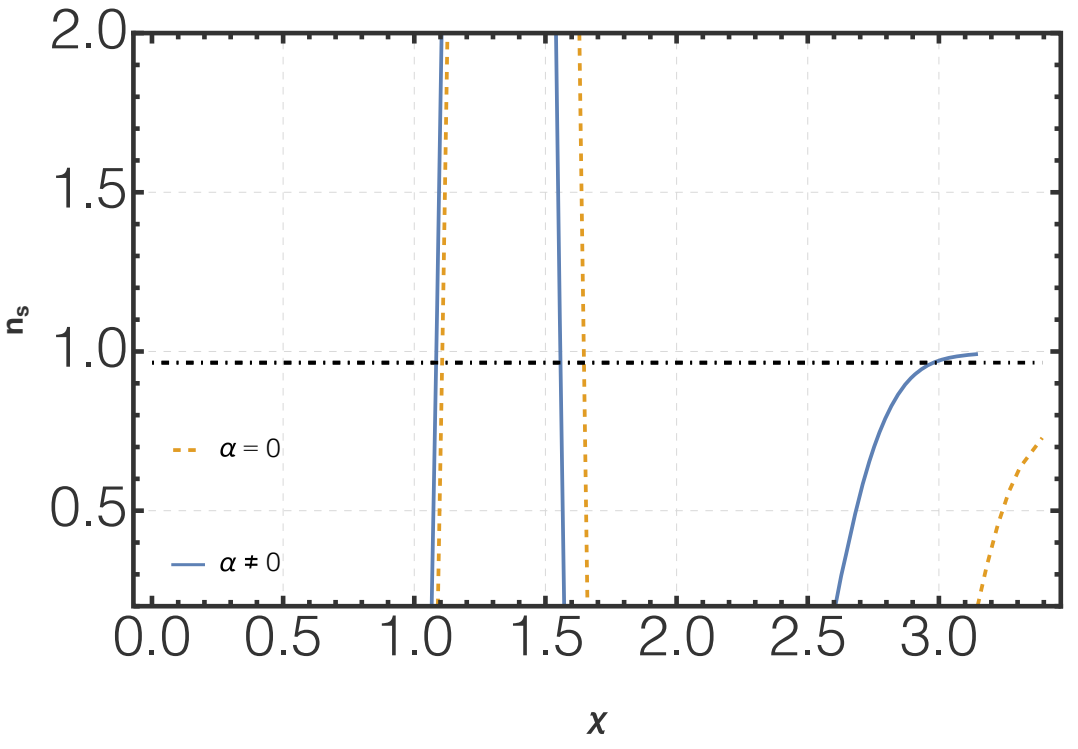

**Fig. 3** Predictions for the scalar spectral index $n_s$ as a function of canonically normalized inflaton field, corresponding to Benchmark 1 of Table 1 with ($\alpha \neq 0$) and without non-minimal coupling ($\alpha = 0$) as shown on the plot legends. The black horizontal dotted-dashed line represents the central value from [80]

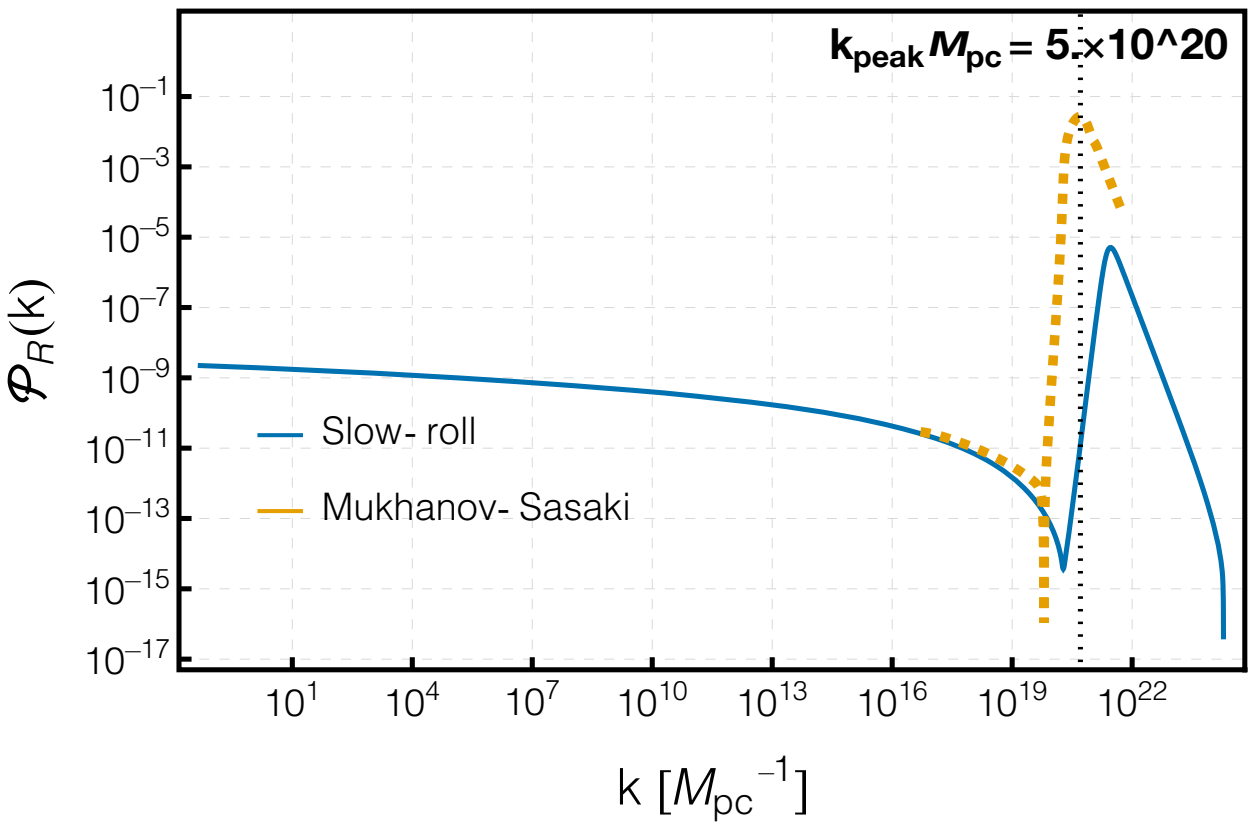

**Fig. 4** Primordial curvature power spectrum as a function of the wave number, shown both in the slow-roll approximation (blue curve) and as obtained from the full solution of the Mukhanov–Sasaki equation (orange dashed line). The potential parameters correspond to Benchmark 1 listed in Table 1. The location of the peak is indicated by a vertical dotted black line and annotated within the figure

## 4 Results

Following the methodology outlined in the previous section, a thorough exploration of the parameter space was performed, combining numerical and manual optimization techniques to maximize the peak amplitude of the power spectrum. Table 1 summarizes five representative benchmark points, each selected independently during a random scan and exhibiting similar phenomenological features. Notably, in all benchmarks, the parameter $f$ is consistently close to $M_{\rm Pl}$ – remaining not strictly sub-Planckian, but also avoiding the extreme trans-Planckian regime often required in standard natural inflation scenarios. This moderation is significant, as exceedingly large trans-Planckian values of $f$ can compromise the stability of the potential through the enhanced impact of higher-dimensional operators.

Another notable outcome of our numerical analysis is that the inflection point consistently lies near the value $\phi_0/f \approx \pi/3$. We find that smaller values of $\phi_0/f$ naturally lead to an increased number of e-folds, as the inflaton must traverse a longer distance to reach the flat region of the potential. Additionally, tuning the peak in the power spectrum becomes significantly more challenging for these smaller inflection point values; the peak amplitude rarely exceeds that at CMB scales. On the rare occasions when it does, the spectrum tends to be broad and peaks at very large scales, but this results in an excessively large number of e-folds, making consistent reheating problematic. Conversely, larger inflection point values are not viable as they render the potential excessively flat near the top of the hill, which also yields an unacceptably large number of e-folds.

Figure 4 underscores the critical importance of accurately estimating the primordial curvature power spectrum by numerically solving the full Mukhanov–Sasaki equation; further technical details are provided in the Appendix C. It is well established that the slow-roll approximation tends to underestimate the peak amplitude and slightly overestimate its location, which can lead to significant inaccuracies in predicting PBH formation and scalar-induced GW signals [14,81]. We have rigorously implemented these considerations to ensure that the resulting power spectrum authentically represents the underlying inflationary physics.

We complement the information presented in Table 1 with Table 2, which shows the predictions for the cosmological parameters, the wave number at the peak of the comoving curvature power spectrum, and the corresponding PBH mass calculated using Eq. (D.1)[7] using the full Mukhanov–Sasaki prediction. We note that Eq. (D.1) assumes PBH formation during radiation domination; this should therefore be regarded as an approximate estimate, especially in the light-PBH regime where the post-inflationary equation of state may differ from radiation domination.

It is noteworthy that nearly all the benchmarks presented were obtained by fixing the spectral index $n_s$ to the lowest value allowed by the Planck results [80]. Higher values of $n_s$ generally lead to a significantly longer period of inflation and shift the peak of the curvature power spectrum to smaller scales, which corresponds to even lighter PBHs. This effect is a well-known challenge: fitting the CMB spectral index within single-field USR inflation models is notoriously difficult [14,81]. Furthermore, the recent data release 6 from the ACT collaboration [20,21] reports a combined spectral index of $n_s = 0.974 \pm 0.003$, which intensifies this tension. Our current scenario is clearly not exempt from these constraints and Benchmarks 1–5 do not survive under the new bounds.

[7] We note that this relation used to compute the PBH masses is only valid if formation occurs during radiation domination while it fails if the dominant species is different [59].

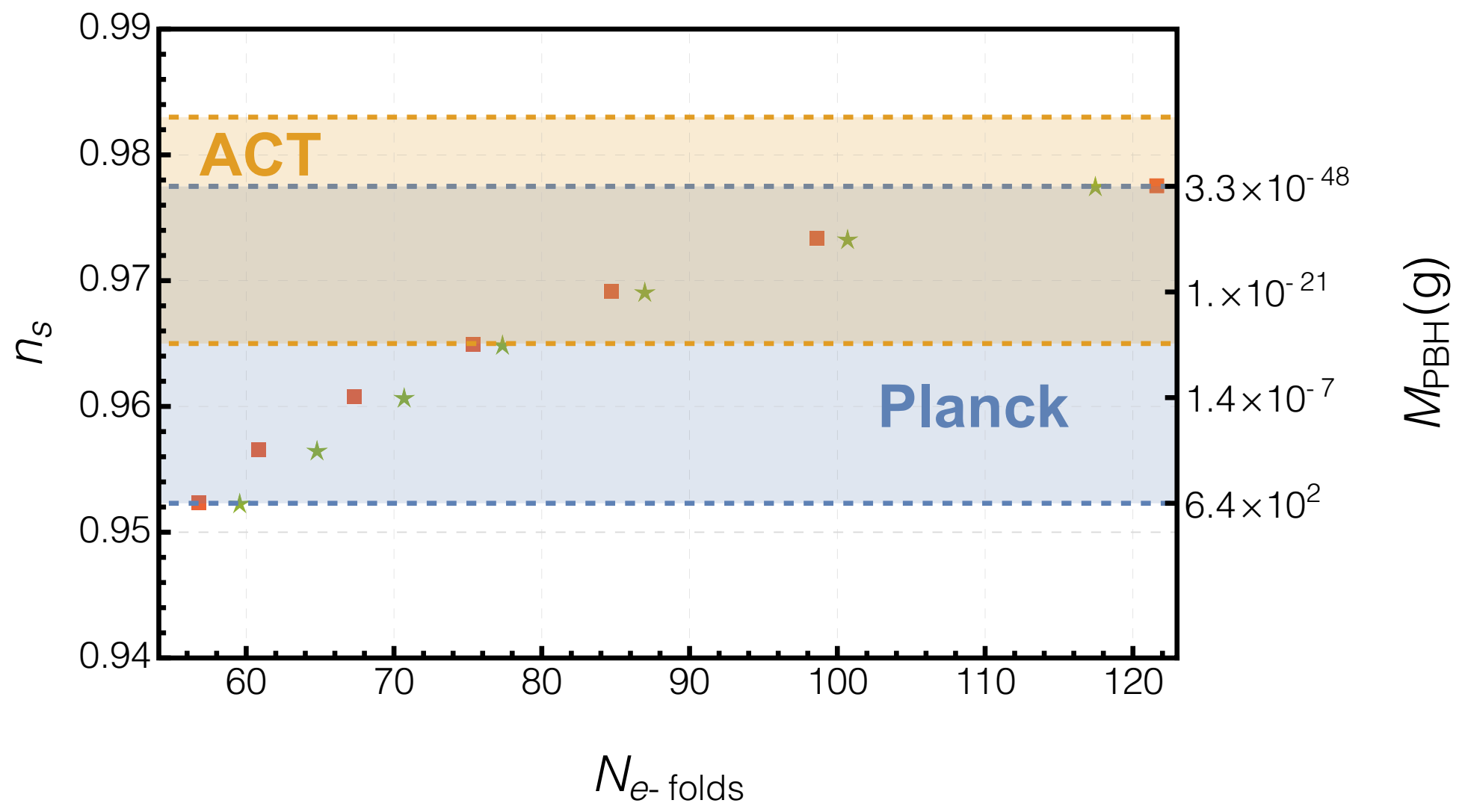

**Fig. 5** Spectral index plotted against the predicted number of e-folds from the time the CMB scale exits the horizon until the end of inflation for Benchmark 1 (stars) and Benchmark 5 (squares). The blue and orange shaded regions represent the $3\sigma$ confidence intervals from Planck [80] and ACT [20], respectively. The secondary y-axis shows the approximate PBH masses at the peak of the power spectrum, calculated using the slow-roll approximation for Benchmark 1; we have verified that Benchmark 5 yields similar PBH mass estimates within the same order of magnitude

Figure 5 illustrates this effect by plotting the spectral index versus the predicted number of e-folds from CMB scale horizon exit until the end of inflation, for Benchmarks 1 and 5 in Table 1. Additionally, using Benchmark 1, we display on the second y-axis, the corresponding PBH masses derived from equation (D.1) within the slow-roll approximation. The qualitative pattern should remain consistent when employing the full Mukhanov–Sasaki formalism. Overall, we conclude that the tightened constraints on $n_s$ represent the strongest probe of this inflationary scenario. If the current ACT results persist, it appears unlikely that the present model will remain a viable theory of inflation, signaling the need for further modifications.

The number of e-folds predicted by our model also warrants scrutiny from the perspective of reheating constraints. To assess the viability of our benchmark points, we compare them with the reheating-matching curves of Ref. [82] in Fig. 6. In this plot, each diagonal line represents the value of $H_k/M_{\rm Pl}$ required by the reheating-matching condition of Ref. [82] for a given constant equation of state $w_{\rm reh}$: a model is compatible with standard cosmology only if it lies to the left of the corresponding curve, since points to the right correspond to modes that re-enter the horizon too late, in conflict with standard cosmological evolution [82]. For $\Delta N_{\rm reh} = 70$, the benchmark points lie between the $w_{\rm reh} = 1/3$ and $w_{\rm reh} = 1/2$ curves.

This reheating analysis reveals a fundamental tension with Big-Bang Nucleosynthesis (BBN). The energy density at the end of reheating is

$$\rho_{rh} = 3M_P^2 H_k^2\, e^{-3(1+w_{\rm reh})\,\Delta N_{\rm reh}}, \tag{4.1}$$

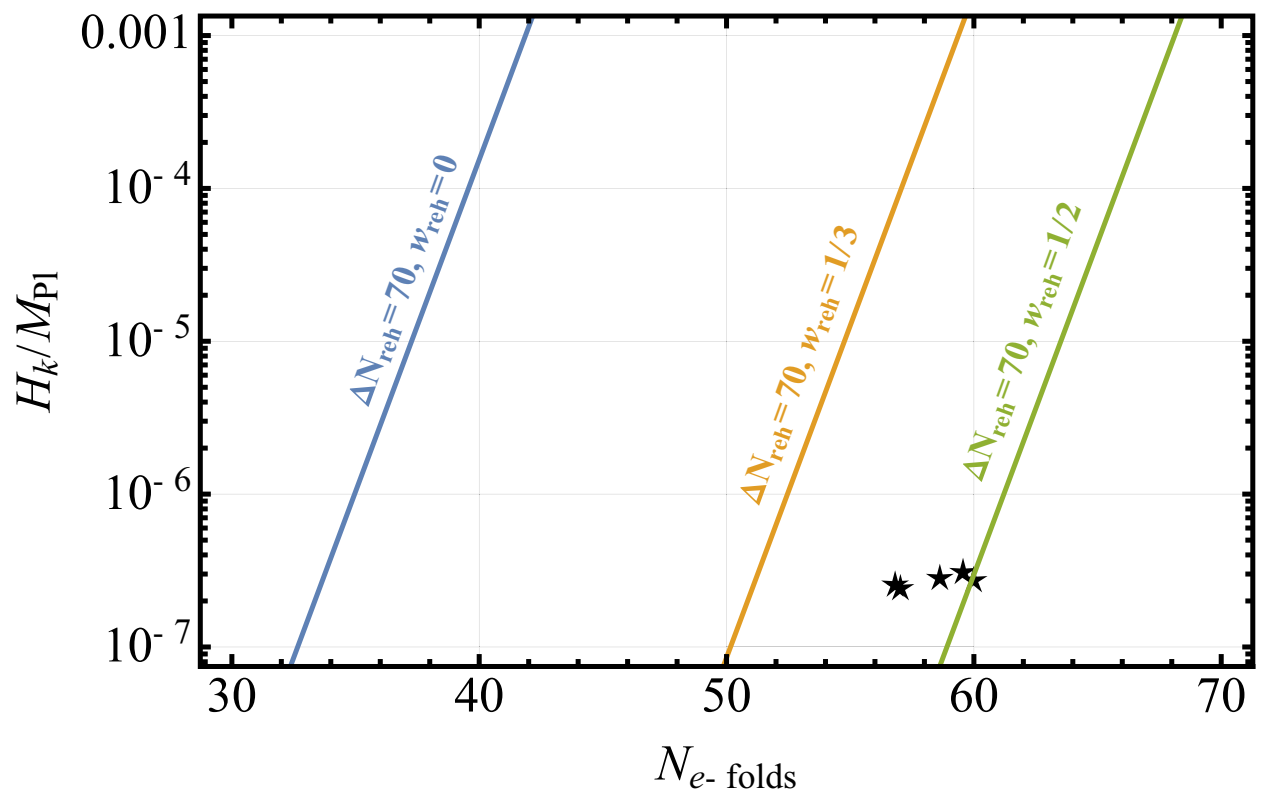

**Fig. 6** Comparison of our benchmark points (stars) with the reheating-matching curves of Ref. [82] for $\Delta N_{\rm reh} = 70$ and different constant equations of state $w_{\rm reh}$. The $y$-axis shows $H_k/M_{\rm Pl}$, the Hubble scale at the end of inflation, and the $x$-axis the number of e-folds between CMB horizon exit and the end of inflation. The region to the right of each curve is excluded; see main text for details. The benchmark points lie between the $w_{\rm reh} = 1/3$ and $w_{\rm reh} = 1/2$ curves, suggesting a relatively stiff effective reheating phase

which for the values $H_k \approx 10^{-7}\, M_P$, $\Delta N_{\rm reh} = 70$, and $w_{\rm reh} = 1/2$ gives $\rho_{rh} \approx 10^{-150}\, M_P^4$. This is catastrophically below the energy density required for BBN to take place in a radiation-dominated universe, $\rho_{\rm BBN} \approx 10^{-85}\, M_P^4$ (corresponding to $T_{\rm BBN} \sim \mathcal{O}(\text{MeV})$). Imposing $\rho_{rh} \geq \rho_{\rm BBN}$ as a hard cosmological constraint, one finds that the maximum allowed number of reheating e-folds at this inflationary scale is

$$\Delta N_{\rm reh}^{\rm max} = \frac{1}{3(1+w_{\rm reh})} \ln\left(\frac{3M_P^2 H_k^2}{\rho_{\rm BBN}}\right) \approx 36\text{–}41 \quad (w_{\rm reh} = 1/3\text{–}1/2), \tag{4.2}$$

roughly a factor of two below the $\Delta N_{\rm reh} = 70$ required by our benchmark points. We have verified that none of the benchmark points studied in this work satisfy this constraint within the standard reheating framework.

This constitutes a genuine tension between the low inflationary scale $H_k \sim 10^{-7}\, M_P$ required by the USR enhancement in the composite pNGB model and the requirement of radiation domination at BBN. We emphasize that this tension is not specific to our model but is a general feature of USR inflationary scenarios producing PBHs in the ultra-light mass range $M_{\rm PBH} \sim 10^3$–$10^5$ g, which necessarily correspond to small-scale perturbations generated at low inflationary scales. Resolving this tension likely requires either a significantly higher inflationary scale or a very rapid energy-transfer mechanism that ends reheating well before BBN, for instance through efficient preheating or instant preheating, as discussed in the standard reheating literature. We therefore present our benchmark points as an exploration of the inflationary and PBH-formation sector of the composite pNGB parameter space, while acknowledging that a fully consistent cosmological history requires addressing this reheating tension as a necessary next step.

Our numerical exploration indicates that the viable benchmarks lead to PBH masses in the range $10^3$ g $\lesssim M_{\rm PBH} \lesssim 10^5$ g, reaching values comparable to everyday civic or even domestic scales. The relevant point here is not the familiar CMB–USR tension itself, but rather the model-dependent outcome of our scan. While other inflationary constructions can realize heavier PBHs without spoiling CMB agreement, we did not find such solutions in the present framework. In this sense, the salient feature is the confinement of viable PBHs to the ultra-light regime within the explored parameter space.

It is worth noting that the PBH mass range identified in this study is a postdiction, as it was not assumed a priori but instead emerged unexpectedly from the underlying model dynamics and analysis. This feature highlights the predictive power and novelty of the proposed mechanism.

Conventionally, PBHs with masses below $10^{14}$ g are expected to have fully evaporated by the present epoch via the standard Hawking radiation mechanism, rendering them unsuitable as dark matter candidates. However, recent theoretical advances [30] have highlighted a memory-burden effect, wherein the black hole evaporation rate is significantly suppressed by multiple powers of its entropy. Considering the exceptionally small PBH masses predicted by our model, we propose that the memory-burden phase initiates early in the evaporation process, significantly suppressing further evaporation and thereby allowing these PBHs to account for the entirety or a substantial fraction of the dark matter.

To quantify this, we compute the PBH abundance using the Press-Schechter formalism, with technical details provided in Appendix D. Figure 7 shows the predicted PBH abundances for Benchmarks 1 through 5. We exclude Benchmark 6 from this analysis because its predicted PBH masses are extremely small, falling well below realistic physical expectations and potentially exceeding the limits of the memory-burden phase framework. Consequently, our focus remains on Benchmarks 1–5, with PBH masses approximately in the range $10^3$ g $\lesssim M_{\rm PBH} \lesssim 10^5$ g. Extending the analysis to even smaller masses would be far more speculative and beyond the scope of this work. We do not include any phenomenological constraint as they are highly dependent on the choices of $k$ and $q$ in Eq. (1.2), see Refs. [31,34]. We stress that this should be regarded as a *leading-order estimate*: primordial non-Gaussianities, unavoidable in USR models, and the use of more refined compaction-function-based statistics can shift the abundance by orders of magnitude (see e.g. [83–86]). The results in this figure are therefore intended as qualitative indicators to identify phenomenologically viable benchmarks, rather than precision predictions.

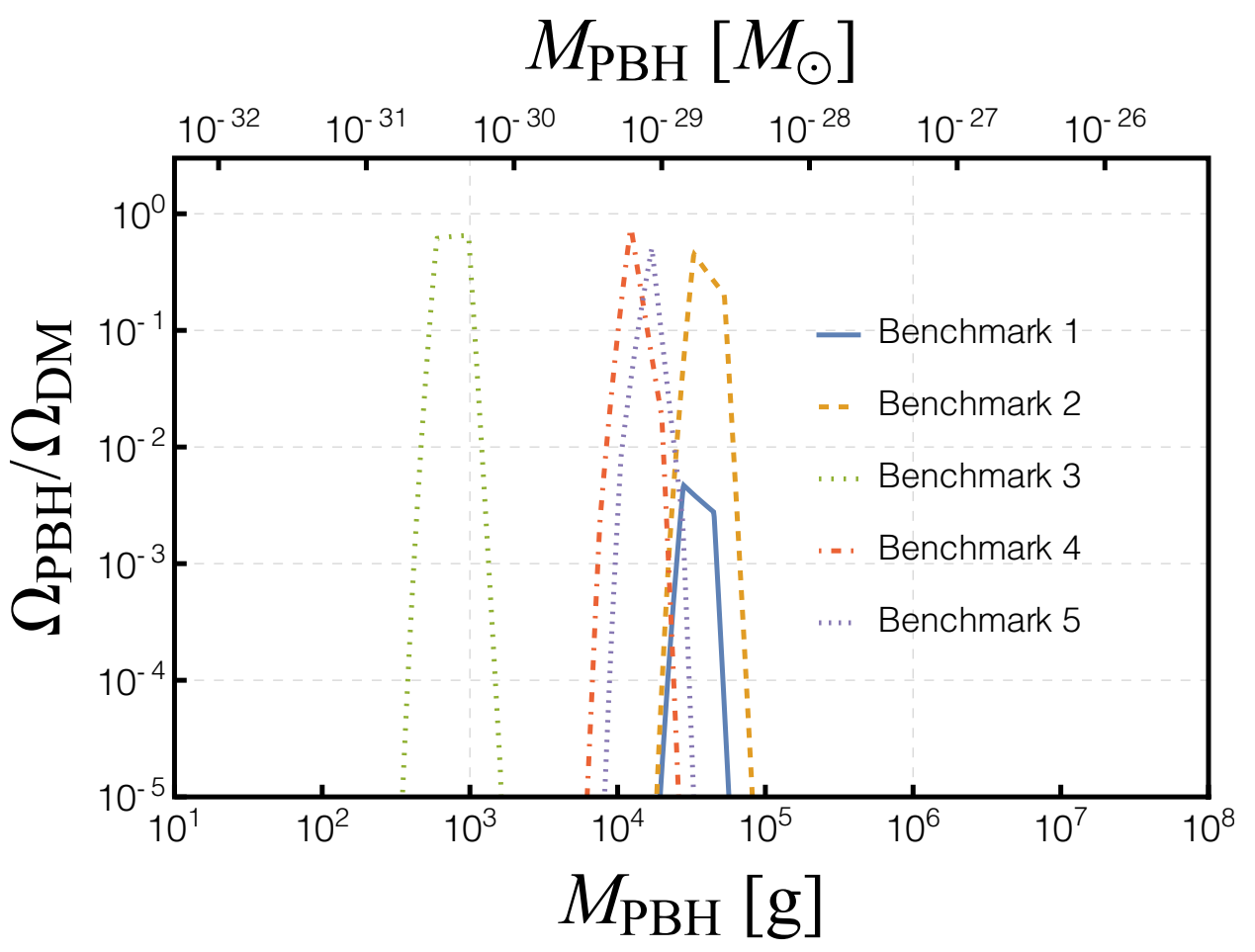

**Fig. 7** PBH abundance for Benchmarks 1–5 in Table 1, computed via the Press–Schechter formalism with Gaussian statistics. This should be regarded as a leading-order estimate; see main text for a discussion of the limitations

We implement the method described in Ref. [45] to compute the GW spectrum and we summarize the most relevant formulas in Appendix E. The resulting GW signals corresponding to the benchmark models listed in Table 1 are shown in Fig. 8, alongside the projected sensitivities of several future detectors[8]: Laser Interferometer Space Antenna (LISA) [47], u-DECIGO [90], Cosmic Explorer (CE) [91], Big Bang Observer (BBO) [92], Einstein Telescope (ET) [93]. The horizontal black lines indicate the current Planck [80] constraints on the effective number of relativistic degrees of freedom,

[8] In the previous version of this paper, the resonant cavity (RC) sensitivity curve [87] was included. However, recent comprehensive analyses [88,89] have challenged the overly optimistic sensitivity estimates previously considered for resonant cavity detectors. We thank Antonio Iovino for bringing these important updates to our attention. Accordingly, we have removed the RC sensitivity curve from our figures in this updated version. Readers should interpret earlier sensitivity projections with caution.

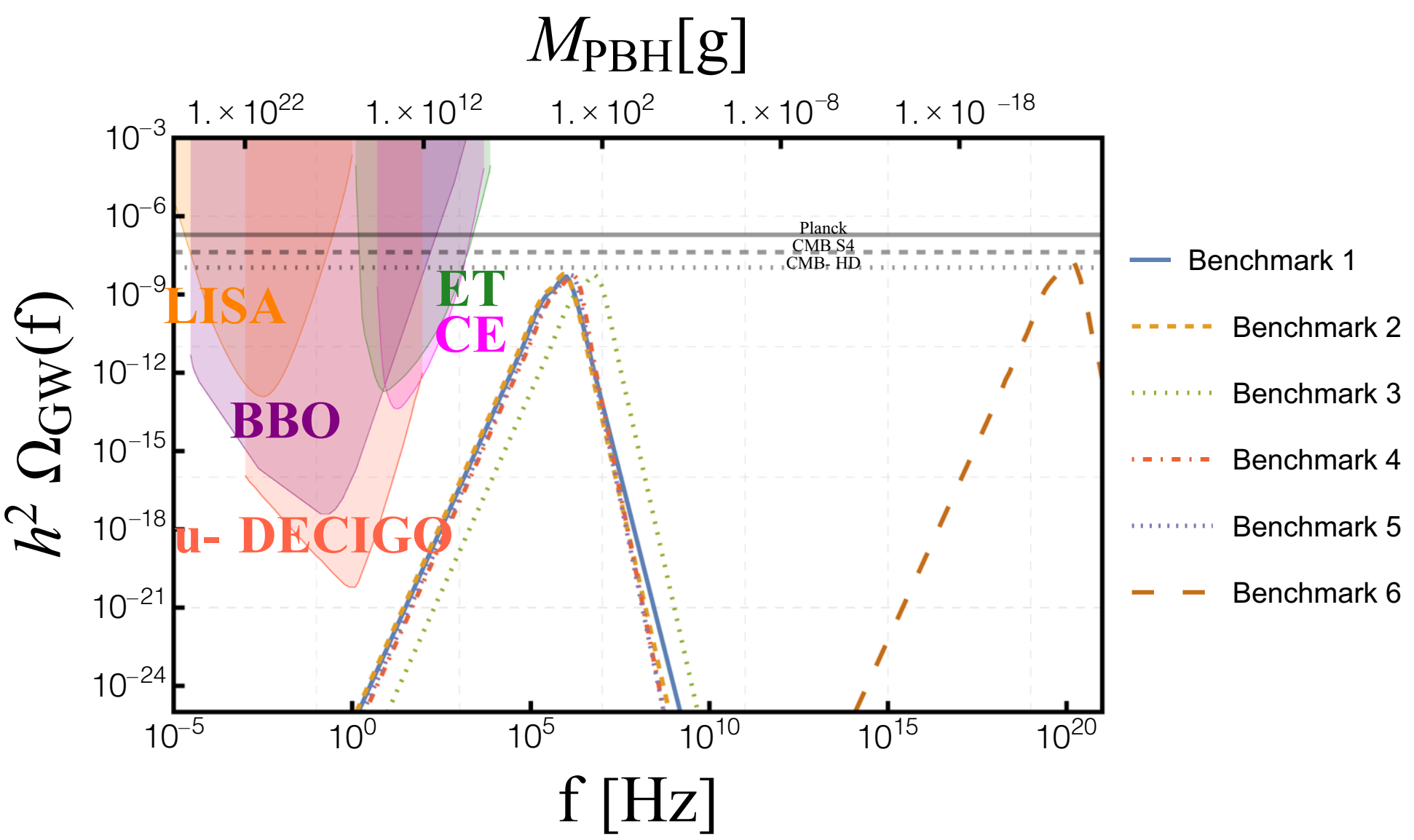

**Fig. 8** GW spectrum of the benchmark models shown in Table 1. We also display experimental sensitivities of various future and proposed detectors (see main text)

along with projected bounds from upcoming experiments CMB-S4 and CMB-HD [94,95].

As evident from the figure, the GW signals peak at high frequencies, residing well beyond the sensitivities of space-based laser interferometers. The peak frequency $f$ of each signal can be approximately estimated using the relation [96]

$$f = \frac{\mathrm{Hz}}{2\pi}\left(\frac{k}{10^{14}\,\mathrm{Mpc}^{-1}}\right),$$

evaluated at $k = k_{\mathrm{peak}}$ which agrees with our numerical results to within approximately 10%. Notably, Benchmark 6 predicts a GW signal at extremely ultra-high frequencies, far exceeding the reach of any currently conceivable detector. However, all predicted signals lie tantalizingly close to the edge of the CMB-HD constraints.

Thus, our scenario introduces a novel benchmark model for ultra-high frequency GW detectors, potentially testable through future measurements of the effective number of relativistic degrees of freedom. At the same time, it remains largely beyond the reach of proposed laser interferometers. This underscores, as often emphasized, the importance of advancing high-frequency GW detection technologies. Potential challenges include the difficulty of probing these frequency bands and the uncertainties in modeling the exact shape and amplitude of the signal near these extremes, which could affect detectability and interpretation.

Finally, we emphasize that the spectral index $n_s$ emerges as the most decisive parameter governing the model's predictions. Essentially, fixing $n_s$ determines the inflaton's position during the slow-roll phase near the potential's hilltop, corresponding to the CMB scales as they left the horizon. In practice, the integration of the inflaton's equation of motion is initialized slightly upstream (further to the right) of where $n_s$ attains the desired value. Larger values of $n_s$ require starting closer to the field boundary on the right, which increases the total number of e-folds, shifts the position of the peak in the scalar power spectrum, and consequently raises the frequency of the induced GW signal. Thus, within our framework, $n_s$ functions not as a prediction but rather as an input parameter from which these critical phenomenological consequences follow.

In its current formulation, the model can accommodate the recent ACT results only by invoking an exceptionally large number of e-folds and extremely small PBH masses. Such small PBHs would necessitate the presence of a memory-burden evaporation phase, which appears excessive to justify their role as dark matter candidates.[9] Furthermore, the associated GW signals peak at ultra- to super-ultra-high frequencies, well beyond the reach of any foreseeable detection technology. Nevertheless, the model remains well motivated from a theoretical standpoint and, to the best of our knowledge, represents the first dedicated exploration of this construction's phenomenology.

## 5 Summary and outlook

In this work, we have investigated a novel single-field inflationary scenario where the inflaton potential originates from a composite sector, analogous to the minimal composite Higgs framework. Our study focused on realizing an USR phase, which generates pronounced peaks in the curvature power spectrum capable of producing PBHs. We have shown that the inclusion of a periodic non-minimal coupling to gravity, carefully matched to the inflaton potential, is essential to achieve sufficient flatness to sustain prolonged inflation and enable the USR phase.

[9] The survival of extremely ultralight primordial black holes until the present epoch may be strongly enhanced in scenarios with extra spatial dimensions or additional particle species, as these factors significantly amplify the memory-burden effect and weaken evaporation constraints; see e.g., [97].

Through a dedicated parameter space scan constrained by CMB observables, we find that the model is tightly restricted, yet highly predictive. Remarkably, the predicted PBH masses span an ultra-light range, $10^3\,\mathrm{g} \lesssim M_{\rm PBH} \lesssim 10^5\,\mathrm{g}$, significantly below conventional astrophysical scales. While such ultra-light PBHs are typically expected to have evaporated via Hawking radiation, the incorporation of memory-burden effects may stabilize these relics, allowing them to survive until today and constitute a significant fraction of dark matter. This speculative but promising avenue links our theoretical predictions to recent theoretical developments regarding black hole evaporation processes.

Assuming that a memory-burdened phase occurs and that ultra-light PBHs constitute a significant fraction of the dark matter, the resulting GW signatures are expected to peak in the ultra-high-frequency regime, far beyond the sensitivity range of conventional laser interferometers. Furthermore, the envisioned CMB-HD experiment, approaches the necessary sensitivity to probe the amplitude of these induced signals. This multi-channel observational frontier motivates continued development of high-frequency GW detection technologies and the design of dedicated experimental strategies bridging cosmology and laboratory-based approaches.

Beyond their potential role as dark matter, the ultra-light PBHs predicted by our model may contribute to baryogenesis mechanisms through evaporation-induced processes [40,42,43], motivating further phenomenological exploration. Nonetheless, like many USR inflationary setups, our model faces tension with the latest high-precision CMB spectral index measurements, notably from ACT, which disfavors the parameter space region favored by our benchmark solutions. This tension marks the spectral index as the most stringent discriminator for this class of models.

In conclusion, our composite pNGB-inspired inflationary model offers a novel and predictive framework that naturally yields ultra-light PBHs and distinct GW signals. Despite its challenges in accommodating the latest CMB constraints, it opens exciting avenues for theoretical refinements and observational probes, enhancing the interplay between particle physics, cosmology and black hole dynamics.

Looking ahead, we plan to explore model extensions incorporating explicit symmetry-breaking operators to relieve fine-tuning and reconcile the scenarios with current cosmological data.

Finally, the computation of the PBH abundance presented in this work relies on the standard Press–Schechter formalism with Gaussian statistics, which provides a leading-order estimate of the abundance. A more refined analysis, incorporating primordial non-Gaussianities inherent to the USR phase [85,86,98] and employing compaction-function-based threshold statistics [83,84,99], is expected to quantitatively modify the predicted abundance for the benchmark points studied here. We leave such a systematic analysis, applied specifically to the composite pNGB model, for future work.

**Acknowledgements** I gratefully acknowledge KTH Royal Institute of Technology for providing institutional support during the completion of this work. I thank the anonymous referee for their valuable assessment and constructive suggestions, which helped improve the clarity and quality of the paper. The author thanks Jonas El Gammal for helpful correspondence. The work of the author is funded by the KMI startup fund and by the KMI/FlaP Young Researchers Grant. I also sincerely thank Antonio Iovino for his helpful feedback and for bringing important recent developments to our attention.

**Data Availability Statement** This manuscript has no associated data. [Authors' comment: No empirical data were generated or analyzed inthis work, as the study is theoretical; therefore, data availability is not applicable.]

**Code Availability Statement** This manuscript has no associated code/software. [Authors' comment: The code was generated solely to perform theoretical computations within this study and is not deposited in a public repository; however, it will be shared with researchers upon reasonable request.]

## A Weyl transformation and canonical normalization

To analyze the inflatonary dynamics, we perform a Weyl transformation to the Einstein frame by redefining the metric as [14]

$$g_{\mu\nu} \to \tilde{g}_{\mu\nu} = F(\phi)^{-1} g_{\mu\nu}.$$

This rescaling removes the non-minimal coupling from the gravitational sector but modifies both the scalar kinetic term and potential. The transformed action becomes

$$\mathcal{S}_{\rm Einstein} = \int d^4x\sqrt{-g}\left[\frac{M_{\rm Pl}^2}{2}R + \frac{K(\phi)}{2}\partial_\mu\phi\partial^\mu\phi - U(\phi)\right], \tag{A.1}$$

where the kinetic function $K(\phi)$ and the Einstein-frame potential $U(\phi)$ are given by

$$K(\phi) = \frac{1}{F(\phi)} + \frac{3M_{\rm Pl}^2}{2}\left(\frac{F'(\phi)}{F(\phi)}\right)^2, \tag{A.2}$$

$$U(\phi) = \frac{V_{\rm inf}(\phi)}{F^2(\phi)}\,. \tag{A.3}$$

The presence of the non-trivial kinetic function generally renders $\phi$ non-canonical. To restore canonical normalization, we define a new field $\chi$ via

$$\frac{d\chi}{d\phi} = \sqrt{K(\phi)}, \quad \chi(0) = 0, \tag{A.4}$$

so that the action simplifies to the standard Einstein-frame form

$$S_{\rm Einstein} = \int d^4x\sqrt{-g}\left[\frac{M_{\rm Pl}^2}{2}R + \frac{1}{2}\partial_\mu\chi\partial^\mu\chi - U(\chi)\right], \tag{A.5}$$

with $U(\chi) = U(\phi(\chi))$.

While analytic expressions for $\chi(\phi)$ exist in some simplified cases (see for instance [18]), the complexity of $F(\phi)$ in our composite-inspired model requires numerical integration for precision. Importantly, the functional form of the non-minimal coupling can substantially flatten the potential in the Einstein frame, enabling prolonged inflation and supporting ultra-slow roll phases necessary for the generation of enhanced curvature perturbations.

## B Inflationay background equations

The slow-roll conditons are written as [14]

$$\epsilon_U = \frac{M_{\rm Pl}^2}{2}\left(\frac{U'}{U}\right)^2, \quad \eta_U = M_{\rm Pl}^2\frac{U''}{U}, \quad \xi_U = M_{\rm Pl}^4\frac{U'U'''}{U^2}, \tag{B.1}$$

where primes denote derivatives with respect to the canonically normalized inflaton field. The scalar spectral index, its running and the tensor-to-scalar ratio are written as

$$n_s \simeq 1 + 2\eta_U - 6\epsilon_U, \quad \alpha_s = -2\xi_U + 16\epsilon_U \\ \eta_U - 24\epsilon_U^2, \quad r \simeq 16\epsilon_U, \tag{B.2}$$

while the overall scale of the potential is fixed by the amplitude of the power spectrum

$$A_s \simeq \left.\frac{U}{24\pi^2\epsilon_U M_{\rm Pl}^4}\right|_{\chi_*} \tag{B.3}$$

We require our models to satisfy the CMB constraints following Planck results [80], namely

$$n_s(\chi_*) = 0.9649 \pm 0.0042, \quad r(\chi_*) < 0.056, \\ \alpha_s(\chi_*) = -0.0045 \pm 0.0067, \quad A_s(\chi_*) = 2.2\times 10^{-9}. \tag{B.4}$$

The equation of motion for the canonically normalized inflaton field in the Einstein frame reads:

$$\frac{d^2\chi}{dN_e^2} + 3\frac{d\chi}{dN_e} - \frac{1}{2M_P^2}\left(\frac{d\chi}{dN_e}\right)^3 + \left(3M_P^2 - \frac{1}{2}\left(\frac{d\chi}{dN_e}\right)^2\right)\frac{d\log U}{d\chi} = 0, \tag{B.5}$$

where we introduced the number of e-folds $dN_e = Hdt$ as our time variable. We solve this equation using slow-roll initial conditions.

The primordial curvature power spectrum at this level can be written as

$$\mathcal{P}_{\mathcal{R}} \simeq \frac{1}{8\pi^2 M_P^2}\frac{H^2}{\epsilon_H}, \tag{B.6}$$

which is the slow-roll approximation.

## C Implementation of the Mukhanov–Sasaki formalism

The accurate computation of the primordial power spectrum of scalar perturbations in inflationary models with features such as near-inflection points or shallow local minima requires going beyond the standard slow-roll approximation. This is achieved by solving the Mukhanov–Sasaki equation numerically, following Ref. [14].

The Mukhanov–Sasaki variable $u_k$ satisfies the mode equation in conformal time $\tau$:

$$\frac{d^2u_k}{d\tau^2} + \left(k^2 - \frac{1}{z}\frac{d^2z}{d\tau^2}\right)u_k = 0, \tag{C.1}$$

where

$$z = \frac{1}{H}\frac{d\chi}{d\tau}, \tag{C.2}$$

with $a$ the scale factor, $\chi$ the inflaton field, and $H$ the Hubble parameter. The comoving curvature perturbation $\mathcal{R}$ is related to $u_k$ by $u_k = -z\mathcal{R}_k$.

For numerical convenience, this equation is rewritten using the number of e-folds $N_e = \ln a$ as time variable:

$$\frac{d^2u_k}{dN_e^2} + (1-\epsilon_H)\frac{du_k}{dN_e}$$

$$+\left[\frac{k^2}{(aH)^2}+(1+\epsilon_H-\eta_H)(\eta_H-2)-\frac{d}{dN_e}(\epsilon_H-\eta_H)\right] u_k=0, \tag{C.3}$$

where the Hubble slow-roll parameters are defined as

$$\epsilon_H=-\frac{\dot{H}}{H^2},\quad \eta_H=\epsilon_H-\frac{1}{2}\frac{d\ln\epsilon_H}{dN_e}. \tag{C.4}$$

Each Fourier mode $u_k$ is evolved starting deep inside the horizon, where it matches the Bunch-Davies vacuum initial conditions:

$$u_k\to\frac{1}{\sqrt{2k}}e^{-ik\tau}.$$

In the e-fold time parameterization, the initial conditions for the real and imaginary parts of $u_k$ and their derivatives are set as

$$\begin{cases}\mathrm{Re}(u_k)=\frac{1}{\sqrt{2k}}, & \mathrm{Im}(u_k)=0,\\ \mathrm{Re}\left(\frac{du_k}{dN_e}\right)=0, & \mathrm{Im}\left(\frac{du_k}{dN_e}\right)=-\sqrt{\frac{k}{2}}\frac{1}{k_i}.\end{cases}$$

The initial e-fold time $N_i$ for each mode is chosen such that the mode is well inside the horizon, with a wavenumber $k_i=k/100$.

After solving the background inflaton dynamics, the Mukhanov–Sasaki equations for each mode are solved numerically and for each of those modes, the power spectrum of curvature perturbations is then extracted at late times (when $k\ll aH$) as

$$\mathcal{P}_{\mathcal{R}}(k)=\frac{k^3}{2\pi^2}\left|\frac{u_k}{z}\right|^2,$$

in practice, for each mode with wave number $k$, we stop the numerical integration when the mode freezes out (when the quantity $\left|\frac{u_k}{z}\right|^2$ goes to a constant value) a few e-folds after exiting the horizon.

This approach allows a precise determination of the enhancement of scalar fluctuations that seed PBH formation, avoiding the inaccuracies of slow-roll approximations which break down near flat features of the potential.

## D Primordial black hole production formulas

PBHs can form in the early Universe from the collapse of large primordial density fluctuations once these fluctuations re-enter the Hubble horizon after inflation. Their formation and abundance can be characterized using the following key formulas, following Ref. [14].

The mass $M$ of a PBH formed at horizon re-entry is typically proportional to the horizon mass at that time:

$$M=\gamma M_H=\gamma\frac{4\pi}{3}\rho H^{-3},$$

where $\rho$ is the energy density, $H$ is the Hubble parameter, and $\gamma\sim 0.2$ parametrizes the efficiency of collapse during radiation domination.

The relation between the comoving wave number $k$ of the perturbation generating the PBH and its mass is given by:

$$M(k)\approx 5\gamma\times 10^{18}\left(\frac{k}{7\times 10^{13}\,\mathrm{Mpc}^{-1}}\right)^{-2}\mathrm{g}. \tag{D.1}$$

Assuming Standard Model relativistic degrees of freedom, this formula connects the scale of enhanced perturbations with the PBH mass.

In the Press–Schechter approach, the fraction of energy density collapsing into PBHs at formation with mass $M$ is:

$$\beta(M)=\int_{\delta_c}^{\infty}\frac{1}{\sqrt{2\pi\sigma^2(M)}}\exp\left(-\frac{\delta^2}{2\sigma^2(M)}\right)d\delta,$$

where $\delta$ is the density contrast, $\delta_c$ is the critical threshold for collapse (typically $\sim 0.45$ during radiation domination), and $\sigma^2(M)$ is the variance of the density contrast smoothed on scale $R=1/k$.

The variance $\sigma^2(M)$ relates to the primordial curvature power spectrum $\mathcal{P}_{\mathcal{R}}(k)$ as:

$$\sigma^2(M(k))=\frac{16}{81}\int\frac{dq}{q}(qR)^4\mathcal{P}_{\mathcal{R}}(q)W^2(qR),$$

with $W(x)=e^{-x^2/2}$ a Gaussian smoothing window.

The present-day fractional energy density in PBHs relative to total dark matter is:

$$\frac{\Omega_{\mathrm{PBH}}(M)}{\Omega_{\mathrm{DM}}}=\left(\frac{\beta(M)}{8\times 10^{-16}}\right)\left(\frac{\gamma}{0.2}\right)^{3/2}\left(\frac{106.75}{g(T_f)}\right)^{1/4}\left(\frac{M}{10^{18}\,\mathrm{g}}\right)^{-1/2},$$

where $g(T_f)$ is the effective relativistic degrees of freedom at PBH formation.

The number of e-folds $\Delta N_e^*$ of inflation elapsed between when a pivot scale $k_*$, e.g. 0.05 $\mathrm{Mpc}^{-1}$, and the PBH formation scale $k_{\mathrm{PBH}}$ exit the horizon relates to PBH mass as:

$$\Delta N_e^*\approx 18.37-\frac{1}{2}\log\left(\frac{M}{M_\odot}\right),$$

where $M_\odot$ is the solar mass. Thus, lighter PBHs correspond to modes exiting the horizon later during inflation.

Importantly, PBH formation requires a sharp enhancement of the primordial power spectrum $\mathcal{P}_{\mathcal{R}}$ at small scales – approximately by seven orders of magnitude relative to the amplitude at CMB scales – to produce an appreciable abundance.

The exponential sensitivity of the mass fraction $\beta(M)$ on the variance $\sigma^2(M)$ and the collapse threshold $\delta_c$ necessitates precise calculation of the primordial spectrum and careful tuning of inflationary model parameters for successful PBH generation. We remark that we neglect the impact of any non-Gaussianity effects as this are known to be negligble in single-field USR inflation [100,101].

**Note on the validity of this approach.** The PBH abundance computed via the Press–Schechter formalism, while standard and widely used in the literature [14], has well-known limitations highlighted by recent work [83,84,99, 102,103]. First, an USR phase unavoidably generates primordial non-Gaussianities [85,86,98], which modify the tail of the PDF of the curvature perturbation and can shift $\beta$ by orders of magnitude. Second, modern analyses employ threshold statistics based on the compaction function, with a shape-dependent threshold and critical-collapse exponent [84,99], rather than a universal fixed threshold applied to the density contrast. Third, the smoothing window function must be chosen consistently with the formation threshold; the Press–Schechter approach with a Gaussian window is known to be internally inconsistent in this regard [84]. The results obtained with this formalism should therefore be understood as *order-of-magnitude estimates*, suitable for identifying promising regions of parameter space but not for precision phenomenology.

## E Computation of the gravitational wave spectrum

This appendix summarizes the computation of the stochastic gravitational wave (GW) background sourced at second order by large scalar perturbations during inflation, following Ref. [45].

The GW spectrum as a function of the comoving wave number is given by

$$\Omega_{\rm GW}(k)h^2 = \Omega_{r,0}h^2 \int_0^{1/\sqrt{3}} \mathrm{d}d \int_{1/\sqrt{3}}^{\infty} \mathrm{d}s\; \Omega_{\rm GW}^{\rm integrand}(d,s,k), \tag{E.1}$$

where the integrand for the GW energy density spectrum is computed as

$$\Omega_{\rm GW}^{\rm integrand}(d,s,k) = \frac{1}{36}\left(\frac{(d^2-\frac{1}{3})(s^2-\frac{1}{3})}{s^2-d^2}\right)^2 \mathcal{P}_\zeta\left(k\frac{\sqrt{3}}{2}(s+d)\right)\mathcal{P}_\zeta\left(k\frac{\sqrt{3}}{2}(s-d)\right)\left[I_c^2(d,s)+I_s^2(d,s)\right] \tag{E.2}$$

and the total GW energy density spectrum $\Omega_{\rm GW}(k)$ is obtained by numerically evaluating the double integral over $s$ and $d$.

In the radiation-dominated era, the kernel functions $I_c(d,s)$ and $I_s(d,s)$ (corresponding to Eqns. D.1 and D.2 in Ref. [45]) are given by

$$\begin{aligned} I_c(d,s) = \frac{288}{(s^2-d^2)^3}\Big[ &(2\cos 1 + (5+d^2)\sin 1)s\sin s \\ &- (2\cos 1 + (5+s^2)\sin 1)d\sin d \\ &+ ((1+d^2)\cos 1 + (5+d^2-2s^2)\sin 1)\cos s \\ &- ((1+s^2)\cos 1 + (5+s^2-2d^2)\sin 1)\cos d \\ &+ \frac{1}{8}(s^2+d^2-2)^2\,(\mathrm{Si}(1+d)+\mathrm{Si}(1-d) \\ &-\mathrm{Si}(1-s) - \mathrm{Si}(1+s) - \pi\Theta(s-1))\Big] \end{aligned} \tag{E.3}$$

and

$$\begin{aligned} I_s(d,s) = \frac{288}{(d^2-s^2)^3}\Big[ &(-2\sin 1 + (5+d^2)\cos 1)s \\ &\sin s - (-2\sin 1 + (5+s^2)\cos 1)d\sin d \\ &+ (-(1+d^2)\sin 1 + (5+d^2-2s^2)\cos 1) \\ &\cos s - (-(1+s^2)\sin 1 + (5+s^2-2d^2)\cos 1)\cos d \\ &+ \frac{1}{8}(s^2+d^2-2)^2\,(\mathrm{Ci}(1+d) \\ &+\mathrm{Ci}(1-d) - \mathrm{Ci}(|1-s|) - \mathrm{Ci}(|1+s|))\Big] \end{aligned} \tag{E.4}$$

where Si and Ci are the sine and cosine integral functions, and $\Theta$ is the Heaviside step function.

The numerical integral for $\Omega_{\rm GW}(k)$ is performed with adaptive quadrature, optimizing integration bounds based on $k$ and the support of $\mathcal{P}_\zeta$. Post-processing includes interpolation and fitting of the spectrum to construct smooth curves for comparison with detector sensitivities and phenomenological analyses. Frequency conversions and normalization factors follow standard conventions (see e.g., Eqs. 4.5, 4.6 in Ref. [49]).

## References

1. A.H. Guth, The inflationary universe: a possible solution to the horizon and flatness problems. Phys. Rev. D **23**, 347 (1981). https://doi.org/10.1103/PhysRevD.23.347
2. K. Freese, J.A. Frieman, A.V. Olinto, Natural inflation with pseudo-Nambu–Goldstone bosons. Phys. Rev. Lett. **65**, 3233 (1990). https://doi.org/10.1103/PhysRevLett.65.3233
3. F.C. Adams, J.R. Bond, K. Freese, J.A. Frieman, A.V. Olinto, Natural inflation: particle physics models, power law spectra for large scale structure, and constraints from COBE. Phys.

Rev. D **47**, 426 (1993). https://doi.org/10.1103/PhysRevD.47.426. arXiv:hep-ph/9207245
4. K. Freese, W.H. Kinney, Natural inflation: consistency with cosmic microwave background observations of Planck and BICEP2. JCAP **03**, 044 (2015). https://doi.org/10.1088/1475-7516/2015/03/044. arXiv:1403.5277
5. D. Croon, V. Sanz, J. Setford, Goldstone inflation. JHEP **10**, 020 (2015). https://doi.org/10.1007/JHEP10(2015)020. arXiv:1503.08097
6. D. Croon, V. Sanz, Saving natural inflation. JCAP **02**, 008 (2015). https://doi.org/10.1088/1475-7516/2015/02/008. arXiv:1411.7809
7. Y. Reyimuaji, X. Zhang, Natural inflation with a nonminimal coupling to gravity. JCAP **03**, 059 (2021). https://doi.org/10.1088/1475-7516/2021/03/059. arXiv:2012.14248
8. N.K. Stein, W.H. Kinney, Natural inflation after Planck 2018. JCAP **01**, 022 (2022). https://doi.org/10.1088/1475-7516/2022/01/022. arXiv:2106.02089
9. A. Salvio, Natural-scalaron inflation. JCAP **10**, 011 (2021). https://doi.org/10.1088/1475-7516/2021/10/011. arXiv:2107.03389
10. A. Salvio, S. Sciusco, (Multi-field) natural inflation and gravitational waves. JCAP **03**, 018 (2024). https://doi.org/10.1088/1475-7516/2024/03/018. arXiv:2311.00741
11. F.B.M. Dos Santos, G. Rodrigues, J.G. Rodrigues, R. de Souza, J.S. Alcaniz, Is natural inflation in agreement with CMB data? JCAP **03**, 038 (2024). https://doi.org/10.1088/1475-7516/2024/03/038. arXiv:2312.12286
12. P. Ivanov, P. Naselsky, I. Novikov, Inflation and primordial black holes as dark matter. Phys. Rev. D **50**, 7173 (1994). https://doi.org/10.1103/PhysRevD.50.7173
13. C. Germani, T. Prokopec, On primordial black holes from an inflection point. Phys. Dark Univ. **18**, 6 (2017). https://doi.org/10.1016/j.dark.2017.09.001. arXiv:1706.04226
14. G. Ballesteros, M. Taoso, Primordial black hole dark matter from single field inflation. Phys. Rev. D **97**, 023501 (2018). https://doi.org/10.1103/PhysRevD.97.023501. arXiv:1709.05565
15. J. Garcia-Bellido, E.R. Morales, Primordial black holes from single field models of inflation. Phys. Dark Univ. **18**, 47 (2017). https://doi.org/10.1016/j.dark.2017.09.007. arXiv:1702.03901
16. B. Carr, K. Kohri, Y. Sendouda, J. Yokoyama, Constraints on primordial black holes. Rep. Prog. Phys. **84**, 116902 (2021). https://doi.org/10.1088/1361-6633/ac1e31. arXiv:2002.12778
17. B. Carr, F. Kuhnel, Primordial black holes as dark matter: recent developments. Annu. Rev. Nucl. Part. Sci. **70**, 355 (2020). https://doi.org/10.1146/annurev-nucl-050520-125911. arXiv:2006.02838
18. G. Ballesteros, J. Rey, M. Taoso, A. Urbano, Primordial black holes as dark matter and gravitational waves from single-field polynomial inflation. JCAP **07**, 025 (2020). https://doi.org/10.1088/1475-7516/2020/07/025. arXiv:2001.08220
19. P.S. Cole, A.D. Gow, C.T. Byrnes, S.P. Patil, Primordial black holes from single-field inflation: a fine-tuning audit. JCAP **08**, 031 (2023). https://doi.org/10.1088/1475-7516/2023/08/031. arXiv:2304.01997
20. ACT Collaboration, The Atacama Cosmology Telescope: DR6 power spectra, likelihoods and ΛCDM parameters. arXiv:2503.14452
21. ACT Collaboration, The Atacama Cosmology Telescope: DR6 constraints on extended cosmological models. arXiv:2503.14454
22. S.W. Hawking, Particle creation by black holes. Commun. Math. Phys. **43**, 199 (1975). https://doi.org/10.1007/BF02345020
23. S.W. Hawking, Black hole explosions. Nature **248**, 30 (1974). https://doi.org/10.1038/248030a0
24. A.M. Green, B.J. Kavanagh, Primordial black holes as a dark matter candidate. J. Phys. G **48**, 043001 (2021). https://doi.org/10.1088/1361-6471/abc534. arXiv:2007.10722
25. A. Escrivà, F. Kuhnel, Y. Tada, Primordial black holes. arXiv:2211.05767
26. M.Y. Khlopov, Primordial black holes. Res. Astron. Astrophys. **10**, 495 (2010). https://doi.org/10.1088/1674-4527/10/6/001. arXiv:0801.0116
27. M. Khlopov, Primordial black hole messenger of dark universe. Symmetry **16**, 1487 (2024). https://doi.org/10.3390/sym16111487
28. G. Dvali, A microscopic model of holography: survival by the burden of memory. arXiv:1810.02336
29. G. Dvali, L. Eisemann, M. Michel, S. Zell, Universe's primordial quantum memories. JCAP **03**, 010 (2019). https://doi.org/10.1088/1475-7516/2019/03/010. arXiv:1812.08749
30. G. Dvali, L. Eisemann, M. Michel, S. Zell, Black hole metamorphosis and stabilization by memory burden. Phys. Rev. D **102**, 103523 (2020). https://doi.org/10.1103/PhysRevD.102.103523. arXiv:2006.00011
31. G. Montefalcone, D. Hooper, K. Freese, C. Kelso, F. Kuhnel, P. Sandick, Does memory burden open a new mass window for primordial black holes as dark matter? arXiv:2503.21005
32. K. Kohri, T. Terada, T.T. Yanagida, Induced gravitational waves probing primordial black hole dark matter with the memory burden effect. Phys. Rev. D **111**, 063543 (2025). https://doi.org/10.1103/PhysRevD.111.063543. arXiv:2409.06365
33. G. Dvali, M. Zantedeschi, S. Zell, Transitioning to memory burden: detectable small primordial black holes as dark matter. arXiv:2503.21740
34. V. Thoss, A. Burkert, K. Kohri, Breakdown of hawking evaporation opens new mass window for primordial black holes as dark matter candidate. Mon. Not. Roy. Astron. Soc. **532**, 451 (2024). https://doi.org/10.1093/mnras/stae1098. arXiv:2402.17823
35. A. Alexandre, G. Dvali, E. Koutsangelas, New mass window for primordial black holes as dark matter from the memory burden effect. Phys. Rev. D **110**, 036004 (2024). https://doi.org/10.1103/PhysRevD.110.036004. arXiv:2402.14069
36. A. Chaudhuri, K. Kohri, V. Thoss, New bounds on memory burdened primordial black holes from big bang nucleosynthesis. arXiv:2506.20717
37. M. Chianese, A. Boccia, F. Iocco, G. Miele, N. Saviano, Light burden of memory: constraining primordial black holes with high-energy neutrinos. Phys. Rev. D **111**, 063036 (2025). https://doi.org/10.1103/PhysRevD.111.063036. arXiv:2410.07604
38. A. Dondarini, G. Marino, P. Panci, M. Zantedeschi, The fast, the slow and the merging: probes of evaporating memory burdened PBHs. arXiv:2506.13861
39. M. Zantedeschi, L. Visinelli, Ultralight black holes as sources of high-energy particles. Phys. Dark Univ. **49**, 102034 (2025). https://doi.org/10.1016/j.dark.2025.102034. arXiv:2410.07037
40. T.C. Gehrman, B. Shams Es Haghi, K. Sinha and T. Xu, Baryogenesis, primordial black holes and MHz–GHz gravitational waves. JCAP **02**, 062 (2023). https://doi.org/10.1088/1475-7516/2023/02/062. arXiv:2211.08431
41. T.C. Gehrman, B. Shams Es Haghi, K. Sinha and T. Xu, The primordial black holes that disappeared: connections to dark matter and MHz-GHz gravitational waves. JCAP **10**, 001 (2023). https://doi.org/10.1088/1475-7516/2023/10/001. arXiv:2304.09194
42. R. Calabrese, M. Chianese, J. Gunn, G. Miele, S. Morisi, N. Saviano, Limits on light primordial black holes from high-scale leptogenesis. Phys. Rev. D **107**, 123537 (2023). https://doi.org/10.1103/PhysRevD.107.123537. arXiv:2305.13369
43. Y.F. Perez-Gonzalez, J. Turner, Assessing the tension between a black hole dominated early universe and leptogenesis. Phys.

Rev. D **104**, 103021 (2021). https://doi.org/10.1103/PhysRevD.104.103021. arXiv:2010.03565
44. S. Datta, A. Ghosal, R. Samanta, Baryogenesis from ultra-light primordial black holes and strong gravitational waves from cosmic strings. JCAP **08**, 021 (2021). https://doi.org/10.1088/1475-7516/2021/08/021. arXiv:2012.14981
45. J.R. Espinosa, D. Racco, A. Riotto, A cosmological signature of the SM Higgs instability: gravitational waves. JCAP **09**, 012 (2018). https://doi.org/10.1088/1475-7516/2018/09/012. arXiv:1804.07732
46. C. Yuan, Q.-G. Huang, A topic review on probing primordial black hole dark matter with scalar induced gravitational waves. iScience **24**, 102860 (2021). https://doi.org/10.1016/j.isci.2021.102860. arXiv:2103.04739
47. P. Amaro-Seoane, H. Audley, S. Babak, J. Baker, E. Barausse, P. Bender, Laser interferometer space antenna. (2017). arXiv preprint arXiv:1702.00786
48. LISA Cosmology Working Group Collaboration, Primordial black holes and their gravitational-wave signatures. arXiv:2310.19857
49. LISA Cosmology Working Group Collaboration, Reconstructing primordial curvature perturbations via scalar-induced gravitational waves with LISA. JCAP **05**, 062 (2025). https://doi.org/10.1088/1475-7516/2025/05/062. arXiv:2501.11320
50. N. Aggarwal et al., Challenges and opportunities of gravitational-wave searches at MHz to GHz frequencies. Living Rev. Relativ. **24**, 4 (2021). https://doi.org/10.1007/s41114-021-00032-5. arXiv:2011.12414
51. G. Laverda, J. Rubio, Higgs-induced gravitational waves: the interplay of non-minimal couplings, kination and top quark mass. JHEP **08**, 203 (2025). https://doi.org/10.1007/JHEP08(2025)203. arXiv:2502.04445
52. K.-Y. Choi, E. Lkhagvadorj, S. Mahapatra, Cosmological origin of the KM3-230213A event and associated gravitational waves. arXiv:2503.22465
53. A. Strumia, G. Landini, Optical gravitational waves as signals of gravitationally-decaying particles. JHEP **04**, 068 (2025). https://doi.org/10.1007/JHEP04(2025)068. arXiv:2501.09794
54. W. Ahmed, A. Ghoshal, U. Zubair, Primordial black holes and second-order gravitational waves in axionlike hybrid inflation. Phys. Rev. D **112**, 063512 (2025). https://doi.org/10.1103/bkkt-4t37. arXiv:2411.00764
55. G. Cacciapaglia, D.Y. Cheong, A. Deandrea, W. Isnard, S.C. Park, Composite hybrid inflation: dilaton and waterfall pions. JCAP **10**, 063 (2023). https://doi.org/10.1088/1475-7516/2023/10/063. arXiv:2307.01852
56. G. Cacciapaglia, D.Y. Cheong, A. Deandrea, W. Isnard, S.C. Park, X. Wang, Composite hybrid inflation: primordial black holes and stochastic gravitational waves. arXiv:2506.06655
57. O. Özsoy, Z. Lalak, Primordial black holes as dark matter and gravitational waves from bumpy axion inflation. JCAP **01**, 040 (2021). https://doi.org/10.1088/1475-7516/2021/01/040. arXiv:2008.07549
58. O. Özsoy, Synthetic gravitational waves from a rolling axion monodromy. JCAP **04**, 040 (2021). https://doi.org/10.1088/1475-7516/2021/04/040. arXiv:2005.10280
59. G. Ballesteros, J. Rey, F. Rompineve, Detuning primordial black hole dark matter with early matter domination and axion monodromy. JCAP **06**, 014 (2020). https://doi.org/10.1088/1475-7516/2020/06/014. arXiv:1912.01638
60. Q. Gao, Y. Gong, Z. Yi, Primordial black holes and secondary gravitational waves from natural inflation. Nucl. Phys. B **969**, 115480 (2021). https://doi.org/10.1016/j.nuclphysb.2021.115480. arXiv:2012.03856
61. J.P.B. Almeida, N. Bernal, D. Bettoni, J. Rubio, Chiral gravitational waves and primordial black holes in UV-protected natural inflation. JCAP **11**, 009 (2020). https://doi.org/10.1088/1475-7516/2020/11/009. arXiv:2007.13776
62. I. Dalianis, A. Kehagias, G. Tringas, Primordial black holes from $\alpha$-attractors. JCAP **01**, 037 (2019). https://doi.org/10.1088/1475-7516/2019/01/037. arXiv:1805.09483
63. M. Drees, Y. Xu, Overshooting, critical Higgs inflation and second order gravitational wave signatures. Eur. Phys. J. C **81**, 182 (2021). https://doi.org/10.1140/epjc/s10052-021-08976-2. arXiv:1905.13581
64. W. Qin, S.R. Geller, S. Balaji, E. McDonough, D.I. Kaiser, Planck constraints and gravitational wave forecasts for primordial black hole dark matter seeded by multifield inflation. Phys. Rev. D **108**, 043508 (2023). https://doi.org/10.1103/PhysRevD.108.043508. arXiv:2303.02168
65. Yogesh, A. Mohammadi, Nonstandard thermal history and formation of primordial black holes and SIGWs in Einstein–Gauss–Bonnet gravity. Astrophys. J. **986**, 35 (2025). https://doi.org/10.3847/1538-4357/adcee5. arXiv:2501.01867
66. K. Agashe, R. Contino, A. Pomarol, The minimal composite Higgs model. Nucl. Phys. B **719**, 165 (2005). https://doi.org/10.1016/j.nuclphysb.2005.04.035. arXiv:hep-ph/0412089
67. R.Z. Ferreira, A. Notari, G. Simeon, Natural inflation with a periodic non-minimal coupling. JCAP **11**, 021 (2018). https://doi.org/10.1088/1475-7516/2018/11/021. arXiv:1806.05511
68. G. Franciolini, P. Pani, Stochastic gravitational-wave background at 3G detectors as a smoking gun for microscopic dark matter relics. Phys. Rev. D **108**, 083527 (2023). https://doi.org/10.1103/PhysRevD.108.083527. arXiv:2304.13576
69. S. Balaji, G. Domènech, G. Franciolini, A. Ganz, J. Tränkle, Probing modified Hawking evaporation with gravitational waves from the primordial black hole dominated universe. JCAP **11**, 026 (2024). https://doi.org/10.1088/1475-7516/2024/11/026. arXiv:2403.14309
70. U. Kumar, Primordial gravitational wave background as a probe of primordial black holes. Phys. Rev. D **112**, 084027 (2025). https://doi.org/10.1103/ymhr-9711. arXiv:2507.10033
71. S.R. Coleman, J. Wess, B. Zumino, Structure of phenomenological Lagrangians. 1. Phys. Rev. **177**, 2239 (1969). https://doi.org/10.1103/PhysRev.177.2239
72. C.G. Callan Jr., S.R. Coleman, J. Wess, B. Zumino, Structure of phenomenological Lagrangians. 2. Phys. Rev. **177**, 2247 (1969). https://doi.org/10.1103/PhysRev.177.2247
73. M. Czerny, F. Takahashi, Multi-natural inflation, in *Phys. Lett. B*. (2014), p.241. https://doi.org/10.1016/j.physletb.2014.04.039 arXiv:1401.5212
74. A. Karam, N. Koivunen, E. Tomberg, V. Vaskonen, H. Veermäe, Anatomy of single-field inflationary models for primordial black holes. JCAP **03**, 013 (2023). https://doi.org/10.1088/1475-7516/2023/03/013. arXiv:2205.13540
75. O. Özsoy, G. Tasinato, Inflation and primordial black holes. Universe **9**, 203 (2023). https://doi.org/10.3390/universe9050203. arXiv:2301.03600
76. K.V. Berghaus, M. Drewes, S. Zell, Warm inflation with the Standard Model. arXiv:2503.18829
77. R. Arya, Formation of primordial black holes from warm inflation. JCAP **09**, 042 (2020). https://doi.org/10.1088/1475-7516/2020/09/042. arXiv:1910.05238
78. M. Bastero-Gil, M.S. Díaz-Blanco, Gravity waves and primordial black holes in scalar warm little inflation. JCAP **12**, 052 (2021). https://doi.org/10.1088/1475-7516/2021/12/052. arXiv:2105.08045
79. M. Correa, M.R. Gangopadhyay, N. Jaman, G.J. Mathews, Primordial black-hole dark matter via warm natural inflation. Phys. Lett. B **835**, 137510 (2022). https://doi.org/10.1016/j.physletb.2022.137510. arXiv:2207.10394

80. Planck Collaboration, Planck 2018 results. X. Constraints on inflation. Astron. Astrophys. **641**, A10 (2020). https://doi.org/10.1051/0004-6361/201833887. arXiv:1807.06211
81. W. Barker, B. Gladwyn, S. Zell, Inflationary and gravitational wave signatures of small primordial black holes as dark matter. Phys. Rev. D **111**, 123033 (2025). https://doi.org/10.1103/4hrv-zfch. arXiv:2410.11948
82. S. Allegrini, L. Del Grosso, A.J. Iovino, A. Urbano, Is the formation of primordial black holes from single-field inflation compatible with standard cosmology? arXiv:2412.14049
83. G. Ferrante, G. Franciolini, A. Iovino Junior, A. Urbano, Primordial non-Gaussianity up to all orders: theoretical aspects and implications for primordial black hole models. Phys. Rev. D **107**, 043520 (2023). https://doi.org/10.1103/PhysRevD.107.043520. arXiv:2211.01728
84. G. Franciolini, A. Iovino Junior, V. Vaskonen, H. Veermae, Recent gravitational wave observation by pulsar timing arrays and primordial black holes: the importance of non-Gaussianities. Phys. Rev. Lett. **131**, 201401 (2023). https://doi.org/10.1103/PhysRevLett.131.201401. arXiv:2306.17149
85. E. Tomberg, Stochastic constant-roll inflation and primordial black holes. Phys. Rev. D **108**, 043502 (2023). https://doi.org/10.1103/PhysRevD.108.043502. arXiv:2304.10903
86. G. Ballesteros, T. Konstandin, A. Pérez Rodríguez, M. Pierre, J. Rey, Non-Gaussian tails without stochastic inflation. JCAP **11**, 013 (2024). https://doi.org/10.1088/1475-7516/2024/11/013. arXiv:2406.02417
87. N. Herman, L. Lehoucq, A. Fúzfa, Electromagnetic antennas for the resonant detection of the stochastic gravitational wave background. Phys. Rev. D **108**, 124009 (2023). https://doi.org/10.1103/PhysRevD.108.124009. arXiv:2203.15668
88. G. Franciolini, A. Maharana, F. Muia, Hunt for light primordial black hole dark matter with ultrahigh-frequency gravitational waves. Phys. Rev. D **106**, 103520 (2022). https://doi.org/10.1103/PhysRevD.106.103520. arXiv:2205.02153
89. N. Aggarwal et al., Challenges and opportunities of gravitational wave searches above 10 kHz. arXiv:2501.11723
90. S. Kuroyanagi, K. Nakayama, J. Yokoyama, Prospects of determination of reheating temperature after inflation by DECIGO. PTEP **2015**, 013E02 (2015). https://doi.org/10.1093/ptep/ptu176. arXiv:1410.6618
91. E.D. Hall, Cosmic explorer: a next-generation ground-based gravitational-wave observatory. Galaxies **10**, 90 (2022). https://doi.org/10.3390/galaxies10040090
92. V. Corbin, N.J. Cornish, Detecting the cosmic gravitational wave background with the big bang observer. Class. Quantum Gravity **23**, 2435 (2006). https://doi.org/10.1088/0264-9381/23/7/014. arXiv:gr-qc/0512039
93. M. Punturo et al., The Einstein Telescope: a third-generation gravitational wave observatory. Class. Quantum Gravity **27**, 194002 (2010). https://doi.org/10.1088/0264-9381/27/19/194002
94. CMB- HD Collaboration, Snowmass2021 CMB-HD White Paper. arXiv:2203.05728
95. CMB- S4 Collaboration, K.N. Abazajian et al., *CMB-S4 Science Book, First Edition* (2016). https://doi.org/10.2172/135204710.2172/1352047. arXiv:1610.02743
96. LISA Cosmology Working Group Collaboration, Gravitational waves from inflation in LISA: reconstruction pipeline and physics interpretation. JCAP **11**, 032 (2024). https://doi.org/10.1088/1475-7516/2024/11/032. arXiv:2407.04356
97. M. Ettengruber, F. Kuhnel, Micro black hole dark matter. arXiv:2506.14871
98. J.H.P. Jackson, H. Assadullahi, A.D. Gow, K. Koyama, V. Vennin, D. Wands, Stochastic inflation beyond slow roll: noise modelling and importance sampling. JCAP **04**, 073 (2025). https://doi.org/10.1088/1475-7516/2025/04/073. arXiv:2410.13683
99. C. Germani, R.K. Sheth, The statistics of primordial black holes in a radiation-dominated universe: recent and new results. Universe **9**, 421 (2023). https://doi.org/10.3390/universe9090421. arXiv:2308.02971
100. V. Atal, C. Germani, The role of non-Gaussianities in primordial black hole formation. Phys. Dark Univ. **24**, 100275 (2019). https://doi.org/10.1016/j.dark.2019.100275. arXiv:1811.07857
101. X.-X. Zeng, Z. Ning, R.-G. Cai, S.-J. Wang, Scalar-induced gravitational waves with non-Gaussianity up to all orders. arXiv:2508.10812
102. I. Musco, V. De Luca, G. Franciolini, A. Riotto, Threshold for primordial black holes. II. A simple analytic prescription. Phys. Rev. D **103**, 063538 (2021). https://doi.org/10.1103/PhysRevD.103.063538. arXiv:2011.03014
103. A.D. Gow, H. Assadullahi, J.H.P. Jackson, K. Koyama, V. Vennin, D. Wands, Non-perturbative non-Gaussianity and primordial black holes. EPL **142**, 49001 (2023). https://doi.org/10.1209/0295-5075/acd417. arXiv:2211.08348